\documentstyle[aaspp4,11pt,psfig]{article}

\begin{document}

\title{The Tully-Fisher Relation as a Measure of Luminosity Evolution: 
A Low Redshift Baseline for Evolving Galaxies\footnote{
Some observations reported in this paper were obtained at the 
Multiple Mirror Telescope Observatory, a facility operated jointly 
by the University of Arizona and the Smithsonian Institution.}}

\author{Elizabeth J. Barton}
\affil{Herzberg Institute of Astrophysics, 
National Research Council of Canada}
\authoraddr{5071 W. Saanich Road,
Victoria, BC, Canada V9E 2E7
(email: Betsy.Barton@nrc.ca)}
\and
\author{Margaret J. Geller}
\affil{Harvard-Smithsonian Center for Astrophysics}
\authoraddr{ Mail Stop 19, 60 Garden St. Cambridge, MA 02138,
        (email: mgeller@cfa.harvard.edu)}
\and
\author{Benjamin C. Bromley}
\affil{Department of Physics, University of Utah}
\authoraddr{201 JFB, Salt Lake City, UT 84112 
(email: bromley@physics.utah.edu)}
\and
\author{Liese van Zee}
\affil{Herzberg Institute of Astrophysics, 
National Research Council of Canada}
\authoraddr{5071 W. Saanich Road,
Victoria, BC, Canada V9E 2E7
(email: Liese.vanZee@nrc.ca)}
\and
\author{Scott J. Kenyon}
\affil{Harvard-Smithsonian Center for Astrophysics}
\authoraddr{ Mail Stop 15, 60 Garden St. Cambridge, MA 02138,
        (email: skenyon@cfa.harvard.edu)}

\begin{abstract}

We use optical rotation curves to 
investigate the $R$-band Tully-Fisher properties of a sample of 
90 spiral galaxies in close pairs, covering a range of 
luminosities, morphological types, and degrees of tidal
distortion.  
The galaxies follow the Tully-Fisher relation
remarkably well, with the exception of
eight distinct $\sim$3$\sigma$ outliers.  
Although most of the outliers
show signs of recent star formation, 
gasdynamical effects are probably
the dominant cause of their
anomalous Tully-Fisher properties.  Four outliers with small
emission line widths have very centrally concentrated 
line emission and truncated rotation curves; the central
emission indicates recent gas infall after a close
galaxy-galaxy pass. These four galaxies may be local 
counterparts to compact, blue galaxies at intermediate
redshift.

The remaining galaxies have a negligible offset from
the reference Tully-Fisher relation, but a shallower
slope (2.6$\sigma$ significance) and a 25\%  
larger scatter. 
We argue that triggered star formation is a significant
contributor to the slope difference.  
We characterize the non-outlier sample with measures of
distortion and star formation
to search for third parameter dependence in the
residuals of the TF relation.
Severe kinematic distortion is the only
significant predictor of TF residuals; this 
distortion is not, however, responsible for the slope difference
from the reference distribution.

Because the outliers are easily removed by sigma clipping,
we conclude that 
even in the presence of some tidal distortion,
detection of moderate 
($\gtrsim 0.5$ magnitudes in rest-frame $R$) luminosity evolution
should be possible with
high-redshift samples the size of this 90-galaxy study.
The slope of the
TF relation, although difficult to measure, is 
as fundamental for quantifying luminosity
evolution as the zero-point offset.

\end{abstract}


\section{Introduction}

Tully-Fisher (1977; TF hereafter) luminosity-linewidth 
studies frequently 
exploit the TF relation as a secondary distance indicator.  
As the fundamental scaling relation for spiral galaxies,
the TF relation also provides constraints on galaxy formation; it is
deeply connected to the processes by which 
disk galaxies form (e.g., Cole et al. 1994; Eisenstein \& Loeb 1996; Navarro \& 
Steinmetz 2000; Mo \& Mao 2000).
By allowing comparisons of galaxies at high $z$ to 
{\it physically similar} systems at the present day,
the TF relation for spirals and the fundamental plane
for spheroid-dominated systems provide crucial anchors
for measuring luminosity evolution directly.

Both the TF and the fundamental plane relations are 
indispensable for establishing a 
complete picture of galaxy evolution.
Kinematic estimates are frequently 
necessary for ascertaining the masses of galaxies.
For example, the faint, compact, blue galaxies may
be small, starbursting galaxies that will eventually fade
or larger galaxies with only moderate amounts of 
star formation (Koo et al. 1994, 1995; Kobulnicky \& Zaritsky 1999).
Only kinematic linewidths that truly reflect the
masses of these galaxies would resolve the degeneracy.
To date, TF studies of galaxies at large redshift 
with resolved rotation curves contain few galaxies and yield
discrepant results; Simard \& Pritchet (1998) find an 
offset of 1.5~--~2  magnitudes, 
while Vogt et al. (1996; 1997) find evidence for only a moderate
offset ($\lesssim 0.4$ magnitudes in $B$).
With the advent of 8-m class telescopes, the size of these 
studies will increase and an understanding of the underlying physics
will be crucial.  
Until now, the primary goal of large TF studies has been 
construction of the tightest 
relation possible for use as a distance indicator; thus
most local TF studies are limited in morphology, 
excluding galaxies with signs of interaction 
and/or tidal distortion (e.g., Rubin et al. 1985; Pierce \& Tully 1992; 
Willick et al. 1995; Willick et al. 1996;
Bureau, Mould, \& Staveley-Smith 1996; Giovanelli et al. 1997a;  
Giovanelli et al. 1997b; Giovanelli et al. 1997c; Courteau 1997; Dale et al. 1997;
Haynes et al. 1999a; Haynes et al. 1999b; Dale et al. 1999; Tully \& Pierce 2000).

With application to high redshift in mind, 
we explore the effects of loosening
the morphological and environmental constraints 
that the largest studies apply.
Pruning of high-$z$ TF samples
is more difficult because interactions are more frequent
and the signatures of interaction are not
so readily observable.  Low surface brightness companions,
minor merger remnants embedded in larger galaxies, 
and faint tidal
features can fall below the detection limits. Seeing effects
obscure both morphological and kinematic distortion.  
The desired measurement of luminosity evolution may become
confused by the effects of tidal distortion on the measurements
of the galaxy parameters. 
Thus, high-redshift studies must rely on 
knowledge of the statistical effects of interactions
gleaned from low-redshift studies.

The TF relation has not yet been extensively
studied as a means of detecting evolution.
We consider the slope of the TF relation and the outliers
to the distribution, in addition to the zero-point offset, 
as fundamental tools for characterizing luminosity evolution 
and morphological evolution.

We explore the TF properties of 90 galaxies
in the Barton, Geller, \& Kenyon (2000a; BGK hereafter)
sample of galaxies in close pairs and n-tuples.  
Because BGK find the
distinct signature of increased star formation due to 
interactions,
the BGK sample provides an ideal testing ground for
the use of the TF relation to detect
moderate luminosity evolution.
We lay the groundwork for future TF studies 
of high-reshift galaxies by: 
(1) quantifying the effects of
moderate distortion on the, intercept, slope and outliers 
of the TF relation, and
(2) setting limits on the ability of pre-
or non-merger interaction to initiate luminosity
evolution off the TF relation at the
current epoch.
Thus, our study complements
searches for TF deviations in low-mass or low-surface brightness
galaxies (Sprayberry et al. 1995; Courteau \& Rix 1999; 
O'Neil et al. 2000; McGaugh et al. 2000),
extreme late-type galaxies (Matthews, van Driel, \& Gallagher 1998),
S0s (Neinstein et al. 1999), and asymmetric galaxies 
(Zaritsky \& Rix 1997), and for 
non-linearity in the relation (Mould, Han, \& Bothun 1989).

In Sec.~2, we describe the sample and the data
reduction and analysis procedures.  We discuss
the TF properties of paired galaxies and identify the
outliers to the TF relation in Sec.~3.  In Sec.~4, we explore the
expected effects of tidal and kinematic distortion and
luminosity evolution; we interpret the outliers in light
of these expectations and identify 4 outliers with similar
systems at intermediate redshift, the compact, blue galaxies with
narrow emission lines.
In Sec.~5, we develop empirical 
measures of distortion and luminosity evolution; we test
these measures for correlations with the TF residuals,
and we examine causes of the differences between
the the pair TF distribution and the reference TF
distribution. We discuss
the implications of our results for high-redshift
evolution studies in Sec.~6 and we conclude in Sec.~7.

\section{The Observations}

In this section, we describe the sample and the data reduction
and analysis procedures.

\subsection{The Sample}

We draw the TF sample from the set of all 
786 galaxies in pairs and n-tuples
in the original CfA2 redshift survey 
with $\Delta D \leq 50$ h$^{-1}$ kpc, $\Delta V \leq 1000$~km/s
and $v \geq 2300$~km/s, where $\Delta V$ is the pair (or n-tuple
neighbor) velocity separation, and $v=cz$ is the apparent recession
velocity.  This TF sample is a subset of the BGK sample;
the full sample is 70\% complete with respect
to all known galaxies in pairs in
their updated CfA2 redshift survey region (Falco et al 1999). 

The sample spans a luminosity range typical of many TF
studies ($-21.74 \leq {\rm M}_{\rm B} \leq -17.68$,
corrected for extinction and assuming 
H$_0 = 70$~km~s$^{-1}$~Mpc$^{-1}$).  The nominal
magnitude limit of the CfA2 redshift survey, 
m$_{\rm B} \approx {\rm m}_{\rm Zw} = 15.5$,
corresponds to an absolute magnitude M$_{\rm B} \sim -21.0$
at the largest redshift in the sample ($z=0.0478$).

In spite of the m$_{\rm Zw} = 15.5$ cutoff, 
the pair sample contains
numerous galaxies with m$_{\rm B} > 15.5$.  
These systems were often
recorded in the CGCG (Zwicky 1961-1968)
with the Zwicky magnitude of the whole system, 
rather than with separate magnitudes for each of the 
galaxies.  Eighteen (of 90) galaxies (20\%) in the TF sample
have m$_{\rm B} > 15.5$, where m$_{\rm B}$ 
is the total $B$ magnitude (see Sec.~2.5).  
The sample is incomplete with respect to these galaxies; we use
the Willick (1994) technique to model the resulting biases
in Sec.~3.

We have different data for different subsamples of
the pairs.  In this paper, we concentrate on the TF
properties of the 90 galaxies for which we have optical rotation
curves, additional nuclear spectra (for all but one), 
and $B$ and $R$ images.
We eliminate galaxies with 
$i < 40^{\circ}$ and with slit misalignment correction
factors $\geq 18.3\%$ [larger than the error in velocity 
width implied by the
1$\sigma$ scatter from the Courteau (1997)
TF distribution, 0.46 magnitudes].  
Sec.~2.4 describes the corrections for inclination and 
misalignment.

\begin{figure}
\centerline{\epsfysize=7in%
\epsffile{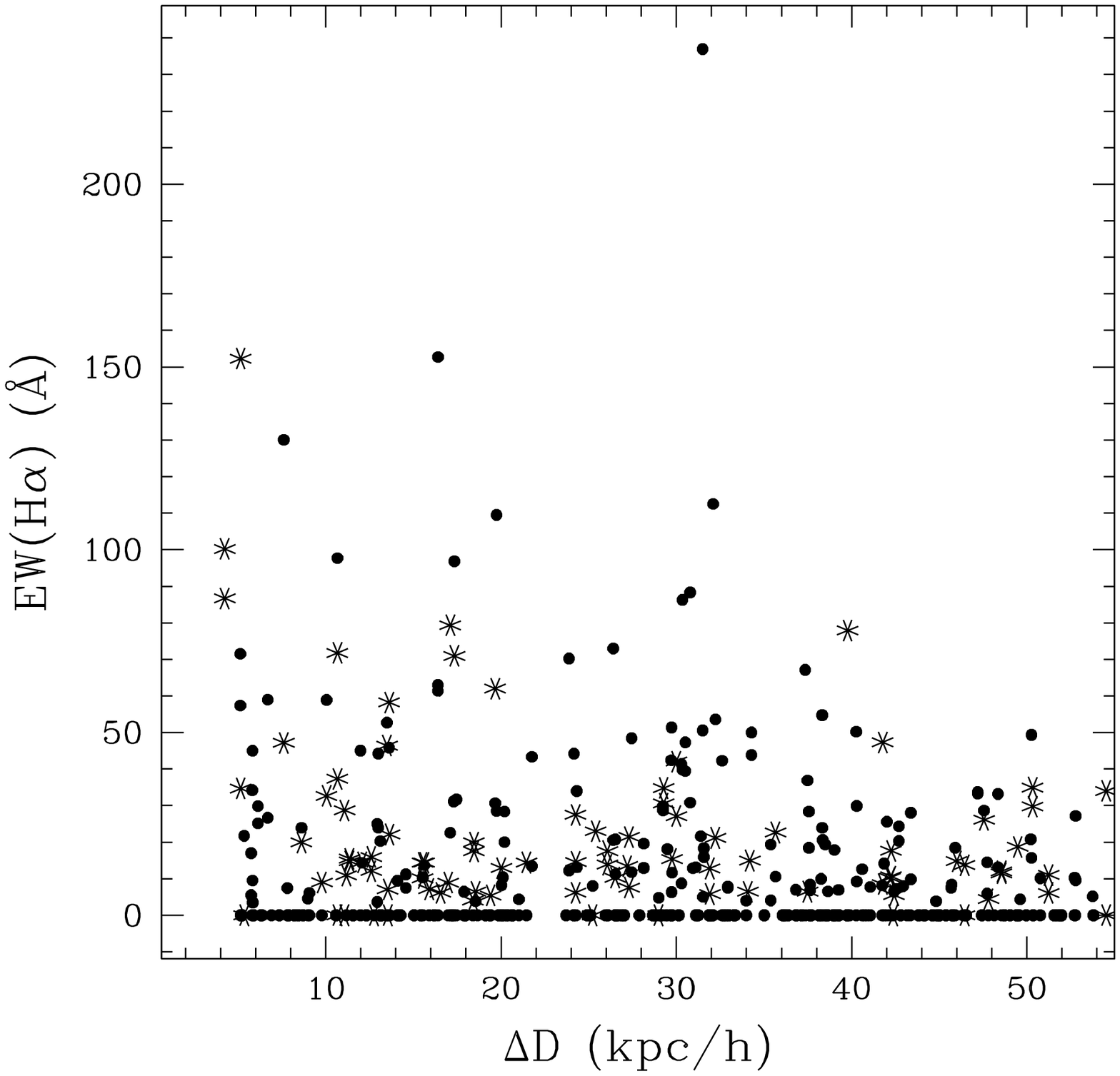}}
\caption{Nuclear EW(H$\alpha$) as a function of $\Delta D$ from Fig.~2 of
Barton, Geller, \& Kenyon (2000).  The stars denote the galaxies
in this study.}
\label{fig:spec}
\end{figure}

We selected systems for rotation curve measurement
based on the availability of H$\alpha$ emission
for the kinematic measurements; this subsample favors
H$\alpha$-emitting galaxies and 
spirals with inclinations $\gtrsim 50^{\circ}$ (see BGK).  
We exclude some of the most distorted galaxies in pairs, typically 
if they no longer appear to have a disk.
Fig.~\ref{fig:spec} shows a modified version of 
Fig.~2 from BGK.  In BGK, the figure shows the 
primary evidence that star formation is triggered by a close
pass --- the EW(H$\alpha$) of
BGK galaxies correlates with pair separation on the sky, $\Delta D$.
Here, starred points show
the placement of the 89 TF target galaxies with nuclear spectra.
For the galaxies 
without significant nuclear H$\alpha$ emission,
we use the disk H$\alpha$ emission 
to measure the rotation curves.

Selection biases play an important role in any
TF study.  However, apart from our deliberate selection
of galaxies in (resolved) pairs or tight systems, 
most TF studies share
our most important biases:  (1) we favor systems with 
substantial line emission, necessary 
to measure optical rotation curves, (2) our magnitude-limited
sample includes intrinsically faint galaxies
only if they are nearby, and
(3) we include only disk galaxies.  Therefore, the effects of
interactions should explain
any significant differences between the TF properties of our sample 
and reference samples.  Sec.~2.4 explores the selection of the
comparison samples.

\subsection{The Emission Line Rotation Curves}

We observed galaxies in the dynamical sample with the Blue Channel
Spectrograph at the Multiple Mirror
Telescope on Mt. Hopkins between November 1996 and February 1998.
For the majority of the galaxies, we used a 1$^{\prime\prime}$ slit, 
with a 1200 lines/mm grating centered at $\sim$6500 \AA.
The spectra cover roughly the wavelength range  5800~--~7200~\AA.
We usually expose for $2 \times 15$ minutes per galaxy.

To reduce the data, 
we employ cross-correlation, which makes simultaneous
use of the major emission lines: [NII] ($\lambda$6548 and
$\lambda$6583), [SII] ($\lambda$6716 and $\lambda$6730), and
H$\alpha$ (see Barton et al. 2000c).
We construct synthetic cross-correlation templates
based on median linewidths and relative heights 
throughout each observing run.  The
technique is straightforward to implement and yields well-defined
errors that enlarge in the presence of multiple velocity
components or low signal-to-noise data.  

\subsection{The Photometry}

We observed the galaxies at the FLWO 48$^{\prime\prime}$ 
telescope on Mt. Hopkins
through either Johnson-Cousins or Harris $B$ and $R$ filters,
for total exposure times of 15 
minutes (usually spread over 3 images) and 5 minutes (with
1 or 2 images),
respectively.  Almost all of the pairs and n-tuples fit
entirely on the Loral $2048\times2048$ CCD, which covers
an $\sim 11' \times 11'$ field, with $\sim$0$^{\prime\prime}$.63 per
pixel when binned by 2.  The observations spanned 
several observing runs between November 1996 and March 1999.

We bias subtract, flatfield, and cosmetically correct 
the data using standard
procedures and the CCDRED tasks in IRAF.

We sky-subtract and measure magnitudes and surface 
brightness profiles
using the GALPHOT package, originally written for surface 
photometry by Freudling (1993),
and updated by N. Grogin (see Grogin \& Geller 1999)
to overlay fitted $B$-band isophotes onto $R$-band images.  
On both the $B$ and $R$ images, we measure the sky level 
in boxes placed around
the galaxy or pair.  We blank out stars on or around the
galaxies using the {\bf imedit} task in the IMAGES package.

With GALPHOT tasks, which make use of several tasks in
the ISOPHOTE package, we fit the remaining (good) data on
the $B$ images with 
isophotal ellipses.  When the signal-to-noise ratio fades in
the outskirts of each galaxy, we fix the center, position angle
and ellipticity based on the inner isophotal ellipses.
Assuming symmetry within the isophotal ellipses (sometimes
not a well-justified assumption) the {\bf sphot}
driver task in GALPHOT constructs a model for 
the ``blanked-out'' data, and
``cleans'' the image with this model.  On rare occasions, when 
the galaxies are very far from axisymmetric, we use 
the {\bf imedit} task to interpolate
locally over stars.  In either case,
{\bf sphot} uses the cleaned image for the final photometry.

{\bf sphot} overlays the $B$ ellipses onto the $R$ image to 
obtain a directly comparable $R$-band surface brightness profile.
It constructs a new $R$ model based on the good available data,
and derives a set of isophotal magnitudes from the cleaned
$R$ image (see Grogin \& Geller 1999 for
details). 

Our study focuses on interacting galaxies; therefore, several
systems are either highly disturbed or contain overlapping galaxies.
Our methods of dealing with these systems
vary from system to
system.  In general, we fit ellipses where possible, 
sometimes smoothing the image for fitting purposes, but we fix
the ellipse parameters if necessary.  When galaxies overlap,
but the individual galaxies are roughly elliptically symmetric,
we derive magnitudes iteratively.  We blank out one of the 
galaxies, fit the second galaxy, subtract the model for the 
second galaxy, and fit the first again.  For most systems, this
procedure is sufficient, but in some cases we continue to 
iterate until the model subtraction appears adequate.
In the worst few cases, when galaxies overlap and are 
not symmetric, we arbitrarily divide the flux between the
galaxies.

We calibrate the images using Landolt (1992) standard fields.
When possible, we calibrate long exposures
directly.  However, we observed most of the systems on mildly to
moderately cloudy nights.  We calibrate these
data using short exposures on photometric nights.  We calculate
photometric offsets with the available stars on the frame.  

Photometric errors vary from system to system.  The most
important factors are: (1) the calibration procedure, 
and (2) galaxy overlap.  A small number of
repeat observations (on non-photometric nights, calibrated
with the same short photometric exposure) suggest
that the photometric
differences between images average $\sim$0.03~magnitudes and 
are $\lesssim 0.05$ magnitudes (with a 
few outliers).
The rms scatter of standard stars on nominally
photometric nights (or half-nights) 
satisfies $\sigma \lesssim 0.023$~magnitudes.
The rms scatter in the calibration from shorter 
exposures falls in the range of $0.015 - 0.07$,
leading to errors in the offsets (sigmas of the mean)
of 0.002~--~0.015 magnitudes.  The additional
errors in the magnitudes due to 
significant overlap of paired galaxies can sometimes
be $\gtrsim 0.1$ magnitudes.
Finally, bright stars nearby on the frame can add errors due
to poor background subtraction.  Sec.~4.3 explores the effects of
some of these errors.

\subsection{The Reference Tully-Fisher Distributions}

We compare our pair sample primarily to the TF study
of Courteau (1997; C97 hereafter), because he derives velocity 
measures from optical rotation curves of comparable quality
and his sample is drawn from regions of quiet Hubble flow, 
outside the cores of rich clusters.  
However, the C97 study uses the Gunn
$r$ filter; our photometry is in $B$ and $R$. 
Therefore, we also consider the  
Pierce \& Tully (1992; PT92 hereafter) and Tully \& Pierce
(2000; TP00 hereafter) radio $B$ and $R$ relations; 
many high-$z$ TF studies use the PT92 $B$ relation 
as a reference distribution.

The sample selection criteria differed for these
three samples.  Unlike our sample which includes
all types of disk galaxies, the C97 sample consists
of unbarred Sb and Sc galaxies in the UGC (see Courteau 1996).
Courteau eliminates ``all peculiar
and interacting galaxies''; he restricts the sample to
inclinations between 55$^{\circ}$ and 75$^{\circ}$,
and m$_{\rm Zw} \leq 15.5$ as for our sample.  
The sample TP92 use for
zero-point fitting includes 15 galaxies in the Local Group, 
the Sculptor group, and the M81 group, with ``normal'' morphology and
low extinction corrections; 6 of these
have reliable independent distance measurements.
They restrict the sample to inclinations $\geq 40^{\circ}$.  
TP92  include 32 galaxies from
the Ursa Major cluster to derive the slope of the TF relation.
Finally, TP00 measure a TF slope from a magnitude-limited
set of galaxies in clusters (both barred and unbarred);
TP00 derive the zero point from a restricted set 
of 24 calibrators.
In summary, our selection criteria differ from the criteria
applied by C97, PT92, and TP00,
but the major difference between our sample 
and the C97, TP92, and TP00 samples is that we 
include only galaxies in close pairs.

\subsection{The Tully-Fisher Parameters}

In this section, we discuss measurements of the TF
parameters.  Our goal is to minimize systematic
differences in measurement techniques between our
pair sample and other TF studies.  Therefore,
we emulate as many of these previously published
techniques as possible.  
We discuss the C97 parameters first.  The adjustments
for comparison to the PT92 and TP00 relations follow.
Table~\ref{tab:measuredparams} lists the set of
parameters we measure.

Like C97, we use ellipse fits to derive corrected total
magnitudes.  However, instead of fitting ellipses in
$r$, we fit ellipses in 
$B$ and then transfer them to $R$.  We
correct the magnitudes from Gunn $r$ to the Johnson-Cousins
system with $r-R=0.354$ (J$\o$rgensen 1994).  Following
C97 we correct the $\mu_{\rm r}=26$ isophotal 
magnitudes to total magnitudes; in some cases,
tidal distortion in the outer isophotes introduces uncertainties
in the extrapolation.
We compute internal reddening following C97 and 
Galactic reddening from n$_{\rm HI}$, the HI column
density.  We compute the absolute magnitudes
assuming the C97 value H$_0 = 70 {\rm\ km\ s}^{-1} {\rm \ Mpc}^{-1}$.
The magnitude errors we compute
include: (1) $\sqrt{N}$ photon noise (from the signal
and the sky), (2) a sky subtraction error of 15\% of
the sky $\sigma$, (3) 0.03 magnitudes added in 
quadrature for photometric calibration (0.05 if we
calibrate off a shorter exposure), (4) 0.1
magnitudes added in quadrature for each case of galaxy overlap or
significant interference from stars on the frame, (5) an
error in Galactic extinction of 20\% of the correction,
(6) an error in internal extinction based on the error in
inclination and, (7) a
distance error from a peculiar velocity assuming 
$\sigma_{\rm v} = \Delta V/2$, where $\sigma_{\rm v}$
is the error in the systemic (galactocentric) velocity of
the galaxy and $\Delta V$ is the velocity separation between the
galaxy and its neighbor (or the member of its n-tuple
with the smallest velocity separation).
The $R$-band total errors range from 0.07 to 0.52 magnitudes.

We usually compute the ellipticity, $\epsilon$, and the
position angle, from the
$\mu_{\rm R} = 24.5$ isophotal ellipse. In 21 cases, 
because of distortion, we use isophotal
ellipses closer to the center of the galaxy, with
$21.5 \leq \mu_{\rm R} \leq 24.0$.  
We estimate the error, $\sigma_{\epsilon}$, from 
the fluctuation in $\epsilon$ as
a function of radius within 5$^{\prime\prime}$ of 
the fiducial isophote, or at the faintest isophote we fit.  
In cases of distortion where we do not fit the ellipses directly,
we assume an error of 0.1 in ellipticity.  
We use C97 Equation~6 to convert from $\epsilon$ to an inclination,
$i$; C97 assume an intrinsic flattening
ratio of q$_0 = 0.18$. 

We compute the disk scale length, R$_{\rm disk}$,
by fitting the surface brightness of the 
outer parts of the galaxy disk with an exponential profile.
In most cases, we follow Courteau's (1996) prescription, computing
$r_{26}$, the radius of the $\mu_{\rm r} = 26$ isophotal ellipse, and
fitting an exponential disk between 0.5$r_{26}$ and 0.9$r_{26}$
(again assuming $r-R=0.354$).
In 4 cases, where distortions affect the outer disks substantially, 
we fit at smaller radii.  In Sec.~4.3, we 
consider the reduced $\chi^2$ of the
fit, $\chi^2_{\rm phot}$, as a measure of
our ability to compute and interpret
R$_{\rm disk}$ parameters accurately.

To measure velocity widths, we follow C97 and fit each 
rotation curve with an empirical fitting function 
(see C97, Equation.~2).  In distorted cases, we 
fit the function anyway; we discuss the effects of
this ``naive'' approach extensively in Secs.~4 and 5.
We measure ${\rm V}_{2.2}$, the velocity width of the galaxy,
by evaluating the difference in the rotation curve model velocities
at radii of $2.15 R_{\rm disk}$ on either side of the galaxy.

Slit misalignment from the major axis 
is an issue for some galaxies; for an optically
and geometrically thin disk, misalignment reduces the measured
velocity width by a factor:
\begin{equation}
f_{\rm align} = 
\left[1+\left(\frac{\tan{m}}{\cos{i}}\right)^2 
\right]^{-1/2},
\label{eqn:misalignment}
\end{equation}
where $m$ is the angle of misalignment between the slit and
the photometric major axis in the plane of the sky
and $i$ is the inclination.
We compute the corrected velocity, V$_{\rm c}$, 
using C97 Eqn.~4 plus the misalignment correction:
\begin{equation}
V_{\rm c} = \frac{{\rm V}_{2.2}}{(1+{\rm z})\sin(i) f_{\rm align}},
\end{equation}
where z is the (galactocentric) redshift of the galaxy.  
Our errors include terms for all of
these factors, assuming the inclination errors described
above, $\sigma_z = 0.00017$, $\sigma_{\rm V2.2} = 8$~km~s$^{-1}$ (see
C97), and $\sigma_m = 2^{\circ}$.  Here, $\sigma_z$,
$\sigma_{\rm V2.2}$, and $\sigma_m$ are the errors in
redshift, V$_{2.2}$, and major axis misalignment angle, respectively.

PT92 and TP00 apply somewhat different prescriptions for their
parameter measurements.  When we compare with their
TF relations, we adjust our Hubble constant and our
measurements of inclination angle and internal extinction (and
the parameters these depend on) to match the prescriptions in 
PT92 (see also Tully \& Fouqu\'{e} 1985) and TP00.  Therefore,
we assign several magnitudes, computed according to the
different distributions, to each galaxy in our sample. 

The biggest systematics, however, probably arise in the
conversions from
optical linewidths to radio 20\%-peak widths, W$_{20}$.  
C97 provides a conversion from V$_{2.2}$ to 50\% widths,
W$_{50}$.  We convert from W$_{50}$ to W$_{20}$ following
Haynes et al. (1999b), and add the turbulence correction
of Tully \& Fouqu\'{e} (1985), to arrive at the
measure of V$_{\rm c}$ which we use when comparing our sample
with the PT92 and TP00 TF relations.

\section{The Tully-Fisher Relation}

\begin{figure}
\centerline{\epsfysize=7in%
\epsffile{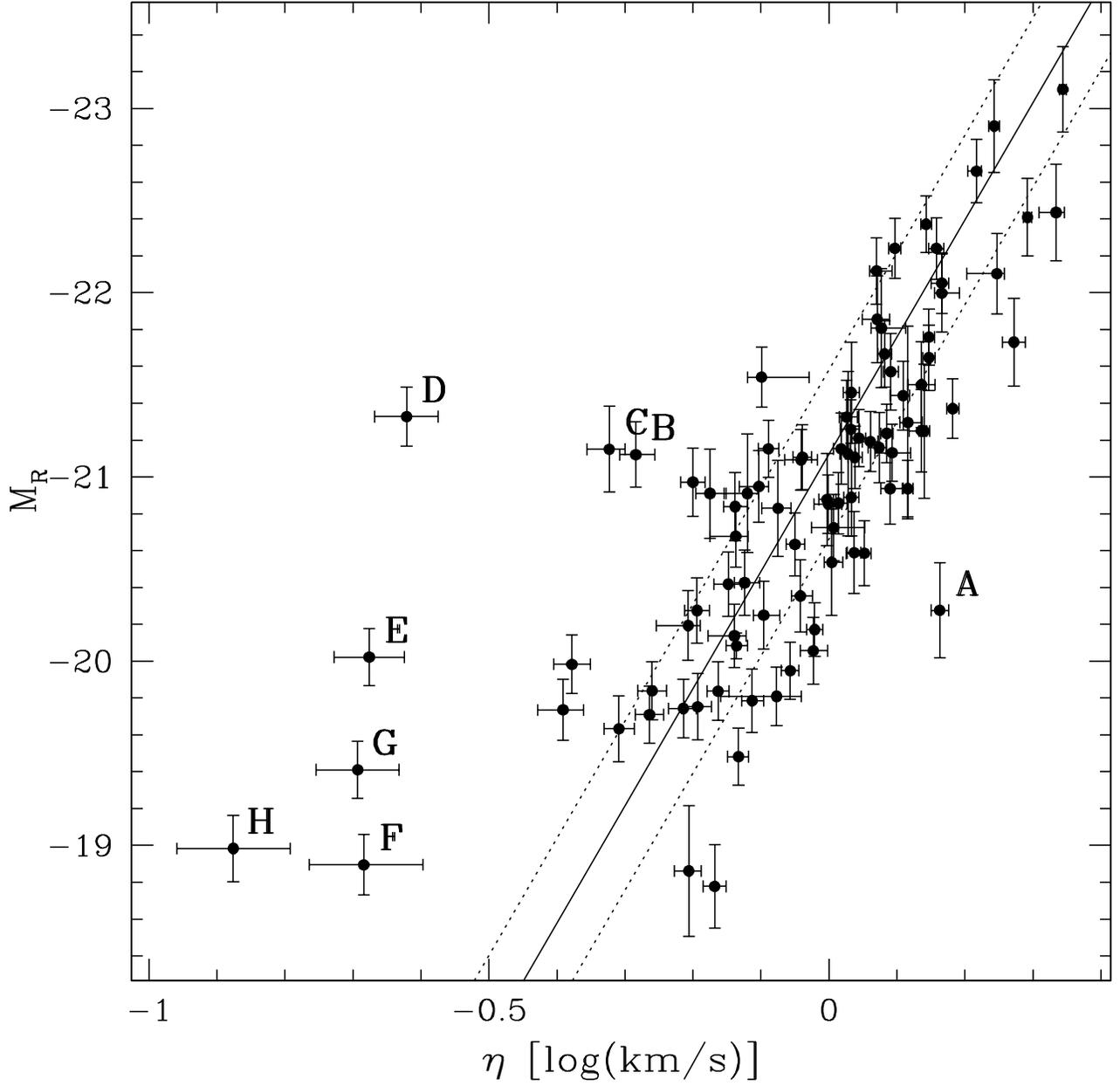}}
\caption{The TF distribution for galaxies in pairs, with
parameters computed with the prescriptions of
Courteau (1997).  The solid line shows the C97 relation
(shifted according to the relation $r-R = 0.354$); the
dotted lines show the C97 1$\sigma$ scatter. }
\label{fig:c97}
\end{figure}

\begin{figure}
\plottwo{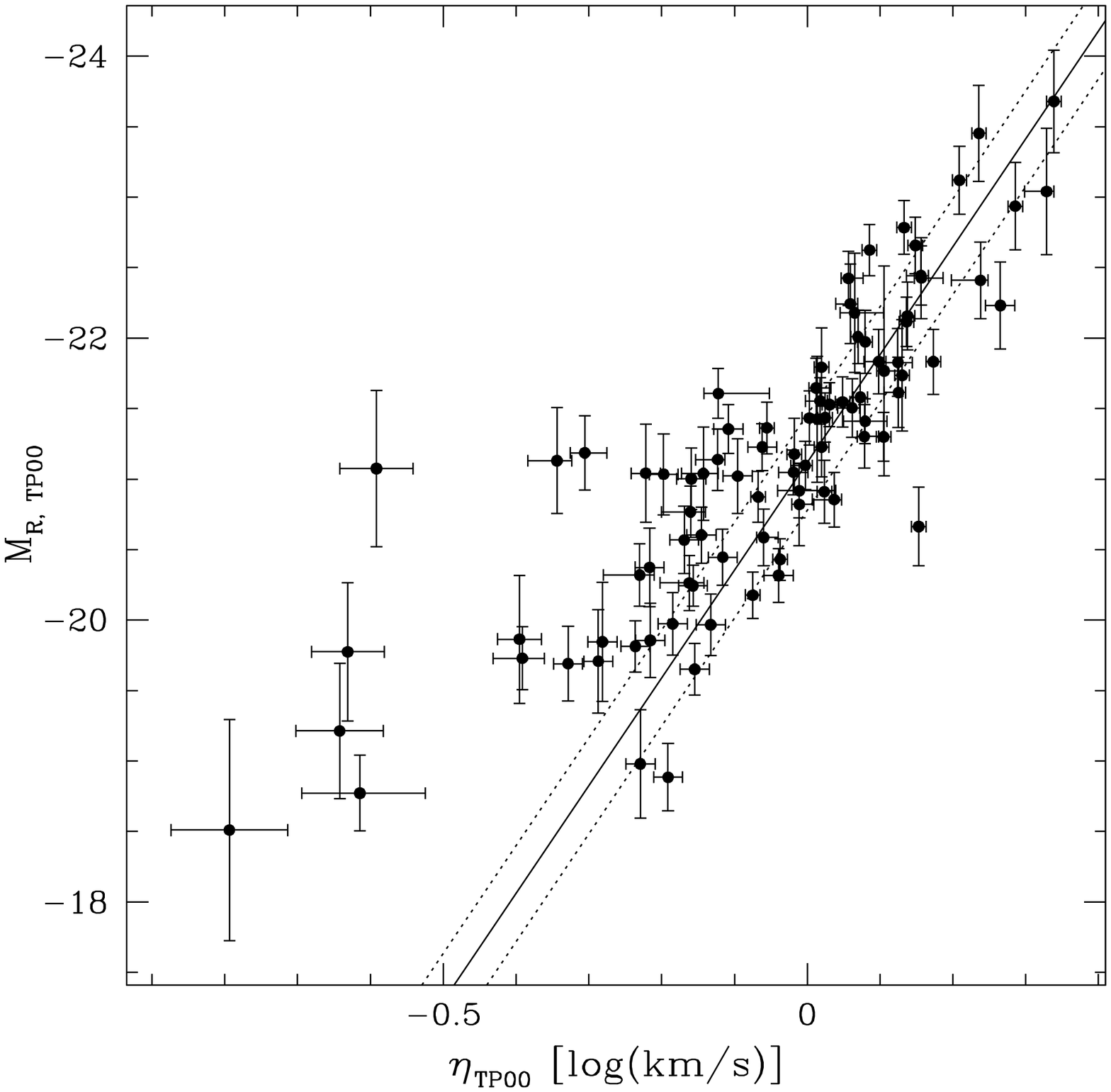}{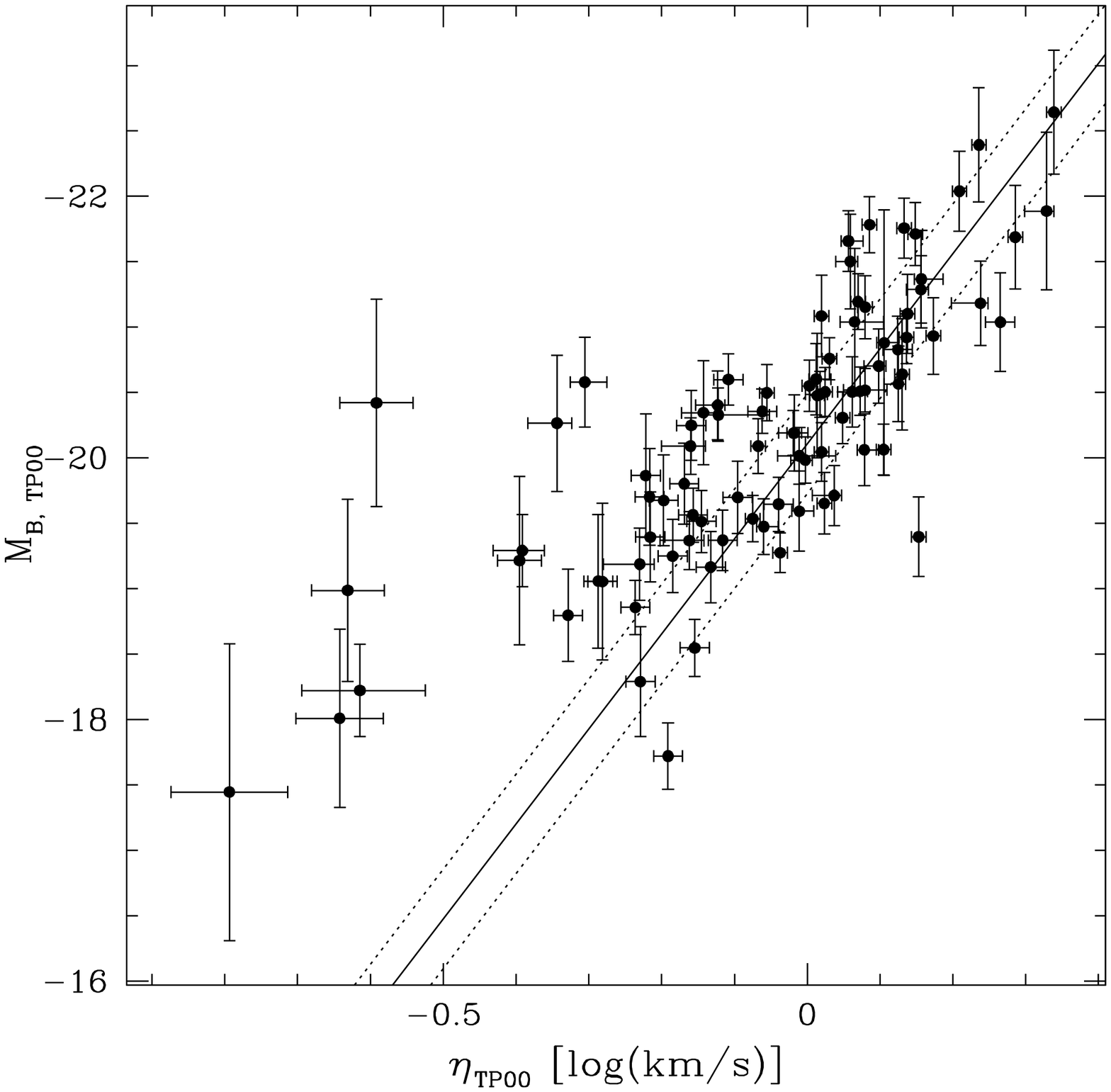}
\caption{The TF relation adjusted to the prescriptions of
Tully \& Pierce (2000) in (a) $R$, and (b) $B$.  In each panel, 
the solid line shows the TP00 relation; the
dotted lines show the TP00 1$\sigma$ scatter.}
\label{fig:tp00}
\end{figure}

In Fig.~\ref{fig:c97} we plot the $R$-band TF relation. We plot
the corrected $R$ magnitude, M$_{\rm R}$, as a function of
$\eta = \log{V_{\rm c}} - 2.5$.
The solid line is Courteau's TF relation (shifted according to the
relation $r-R=0.354$ from J$\o$rgensen 1994); the dotted lines
illustrate his 0.46-magnitude scatter.  We 
show the TP00 (cluster/group galaxy) 
relations in Fig.~\ref{fig:tp00}.

Fig.~\ref{fig:c97} shows that the galaxies lie on 
a TF relation similar to
the (optical) relation of ``field''
galaxies, with only a few exceptions near the faint end.  
As a first measure of the agreement with the reference TF
relations, we list the offset and scatter in 
Table~\ref{tab:tf}.  We compute these measures with a prescription 
similar to that of Vogt et al. (1997), 
by fitting for the Tully-Fisher offset while
keeping the slope fixed to the slope of the reference 
distribution. 
We quote the sigma-of-the-mean as the error in the offset, 
and the rms value of the
difference between the measured magnitude 
and the expected magnitude,
offset by $\Delta_{\rm TF}$, as the scatter.
We detect no net offset from the C97 TF distribution with
this simple measure; the
scatter increases substantially, however, from the C97 value
of 0.46 magnitudes to 1.0 magnitude here.

The radio TF relations present a somewhat different picture.
We find substantial deviations, $\Delta_{\rm TF}$, from the
TP00 distribution (0.46~--~0.50 magnitudes) and even larger
deviations from the PT92 distribution (0.75~--~0.87 magnitudes).  
In principle, we eliminate simple systematic errors which originate
from different measurement techniques by altering our prescriptions for
computing the TF parameters.  Barring errors in the C97 radio
to optical conversion,
the different offsets between our sample and the reference
distributions probably reflect {\it real} discrepancies among
the reference distributions.
Wherever the errors lie, the discrepancies between the C97, PT92, and TP00 
TF distributions have important implications for
any study which compares the TF properties of a set of galaxies
to a TF reference distribution.
In particular, the discrepancies may be responsible for the 
0.4--magnitude offset for high-$z$ galaxies from 
Vogt et al. (1997), who compare to the PT92 $B$-band distribution.
(Vogt et al. are careful to note that their measured offset is
an upper limit.)

In the following three sections, we develop a more rigorous 
test for true departures from the C97 reference TF distribution.  
In Sec.~3.1 we apply the
Willick (1994) correction for luminosity bias to determine
the TF slope and offset for each sample.  We describe a Monte
Carlo test in Sec.~3.2 which we use to measure the significance
of offsets.  In Sec.~3.3, we use the Monte Carlo simulations
to argue that several of the apparently wayward points 
in Fig.~\ref{fig:c97}
are true, non-Gaussian outliers; we eliminate 8 of these outliers
and recompute the sample TF properties.

\subsection{Slope and Zero-Point Offsets from a Bias-Correcting
Analysis}

Figs.~\ref{fig:c97} and \ref{fig:tp00} illustrate 
another shortcoming of computing fixed-slope TF 
offsets from reference distributions.  In both 
figures, the TF relations show some evidence
for slope differences.  To describe the TF relations, 
we require a more complete description of
the TF properties of our sample.  Luminosity
biases introduced by restriction to a magnitude-limited
sample, coupled with the intrinsic dispersion in the TF
relation, strongly affect the measured slope of the
TF distribution (e.g., Willick 1994). At a given velocity width
towards the low-mass end of the distribution, only the 
most luminous galaxies appear in the TF sample.  This
effect results in a bias towards a shallow measured slope 
for the ``forward''
TF relation (where linewidth is the independent variable).

Here, we analyze the $R$-band TF relations of the pair
sample and the C97 sample using the luminosity bias
correction technique of Willick (1994).  For the C97
sample, we eliminate 5 galaxies which do not meet
the original selection criteria; 2 are not in
the UGC, 2 have UGC diameters $< 1^{\prime}$ and 
two have m$_{\rm Zw} > 15.5$.
We omit the correction for diameter limit bias in the C97
sample.  This correction increases the measured differences 
between the C97 TF distribution and the pair relation
by a small amount ($\lesssim 2\%$ in the slope and offset
and $\sim$3\% in the scatter).

We apply the luminosity bias correction closely 
following Willick (1994).  The first step
is to relate the selection parameter, in this case the Zwicky 
magnitude, m$_{\rm Zw}$, to the TF parameters $M$ and $\eta$.
We use the functional form adopted in Willick (1994), 
m$_{\rm Zw} = {\rm a} + {\rm b} \eta + $~cm, where a, b, and c are
the coefficients we fit and m is the relevant apparent
magnitude ($R$ for the pairs, and $r$ for the
C97 sample).  We use the bias-correcting 
technique outlined in Sec.~3 of Willick (1994) to fit for a, b, and c,
as well as for $\sigma_{\rm M,Zw}$, the scatter in the relationship. 
Table~\ref{tab:mzrel} lists the results.  

Next, we correct for the bias introduced because
only galaxies with m$_{\rm Zw} \leq 15.5$ appear in the
samples (C97 and our sample).  The technique is
iterative. The steps are: (1) estimate the TF parameters, 
(2) using the relationship between the TF
parameters and the selection parameter, m$_{\rm Zw}$,
compute the expectation 
value of the measured apparent magnitude, m, as a function of 
$\eta$, and correct each point for the difference between 
this expectation value and the ``true'' value of m from the 
TF relationship found in (1); (3) re-compute the TF relationship,
and (4) iterate on steps (2) and (3) until the TF 
slope and offset converge.

\begin{figure}
\centerline{\epsfxsize=6.5in%
\epsffile{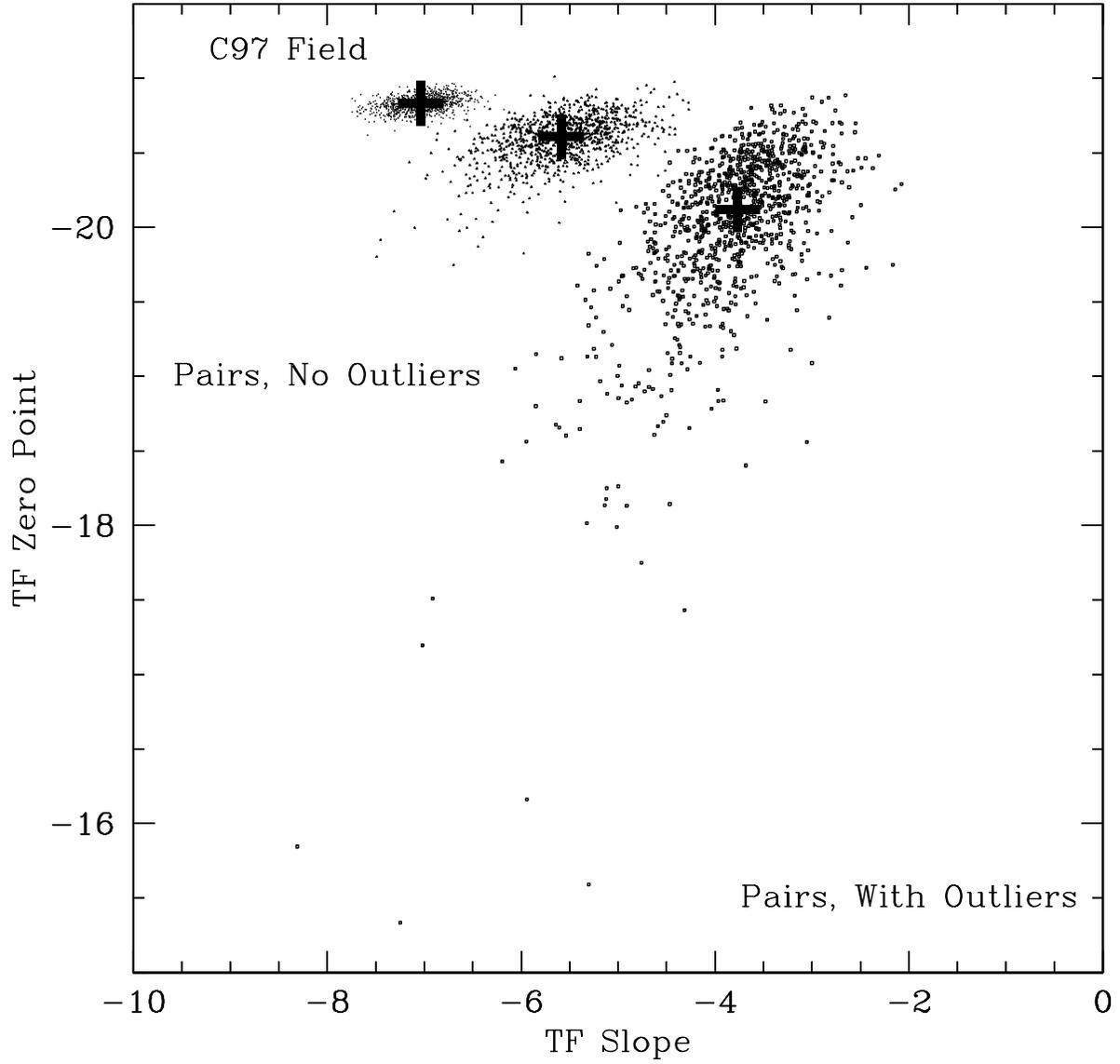}}
\caption{TF fits to full sample, our sample without the 8 outliers, 
and the C97 data (shifted via $r-R =0.354$).  The
crosses are the solutions, the smaller points are the results from
the Monte Carlo simulations, indicating the confidence intervals.}
\label{fig:tfparams}
\end{figure}

In Table~\ref{tab:tfparams} we list the TF parameter fits 
resulting from this technique
for the C97 data and for our galaxies in pairs.  
Because of this bias correction, the slope we measure
for the C97 data, -7.03,  is steeper than the slope
C97 compute, -6.36.
Fig.~\ref{fig:tfparams} shows the solutions (crosses).
The points represent the Monte Carlo results below.

\subsection{Computing Confidence Intervals with a Monte Carlo 
Simulation}

We use a Monte Carlo simulation to estimate the confidence 
levels appropriate for the TF parameter solution.  
We draw model datasets from the best-fit 
TF relation (based on the analysis in Sec.~3.1), 
subject the ``data'' to the selection effects relevant to each
sample, and use the Willick (1994) technique to
find the best-fit TF parameters of the simulated data.  

To model the TF relation for the pair data, 
we draw absolute magnitudes, $M$,
from the void galaxy LF of Grogin \& Geller (1999) 
($M_{\star} = -20.4$, $\alpha_{\rm LF} = -1.17$)
because Grogin \& Geller derive their LF from data of 
very similar quality based on
a survey of similar depth to our pair sample. 
The Grogin \& Geller LF is consistent
with the Century Survey LF (Geller et al. 1997).

After drawing $M$ from the LF, we convert to $\eta$ using the
input TF distribution, the solution found via the
Willick technique.  We then mimic scatter in the forward
TF distribution by adding a random number to $M$ generated from
a Gaussian distribution with dispersion $\sigma_{\rm TF}$, where
$\sigma_{\rm TF}$ is the scatter in the TF
distribution measured from the 
data.  We compute the apparent magnitude, m, 
from $M$ by assigning a
randomly drawn velocity appropriate
for a fixed-solid-angle survey with 
2300~km~s$^{-1} \leq cz \leq 15000$~km~s$^{-1}$; 
we draw a value of m$_{\rm Zw}$ for each point from the
relationship m$_{\rm Zw} = {\rm a} + {\rm b} \eta + $~c\,m,
derived from the data, with scatter $\sigma_{\rm m, Zw}$.
Finally, we eliminate points with m$_{\rm B} > 15.5$ to
mimic the selection bias.  

For the pair study, we draw ``datasets'' of
the appropriate number of galaxies (i.e., 90 
for the pair sample and 279 for the C97 sample).  We use the 
Willick (1994) procedure to find 
the best-fit TF solution for each ``dataset''.
Fig.~\ref{fig:tfparams} shows the results for 1000 simulations of
each distribution.  The results are centered on the input solution,
demonstrating that the Willick (1994) technique 
properly accounts for luminosity bias.

\subsection{Identifying the Outliers to the Pair TF Distribution}

\begin{figure}
\centerline{\epsfxsize=6.5in%
\epsffile{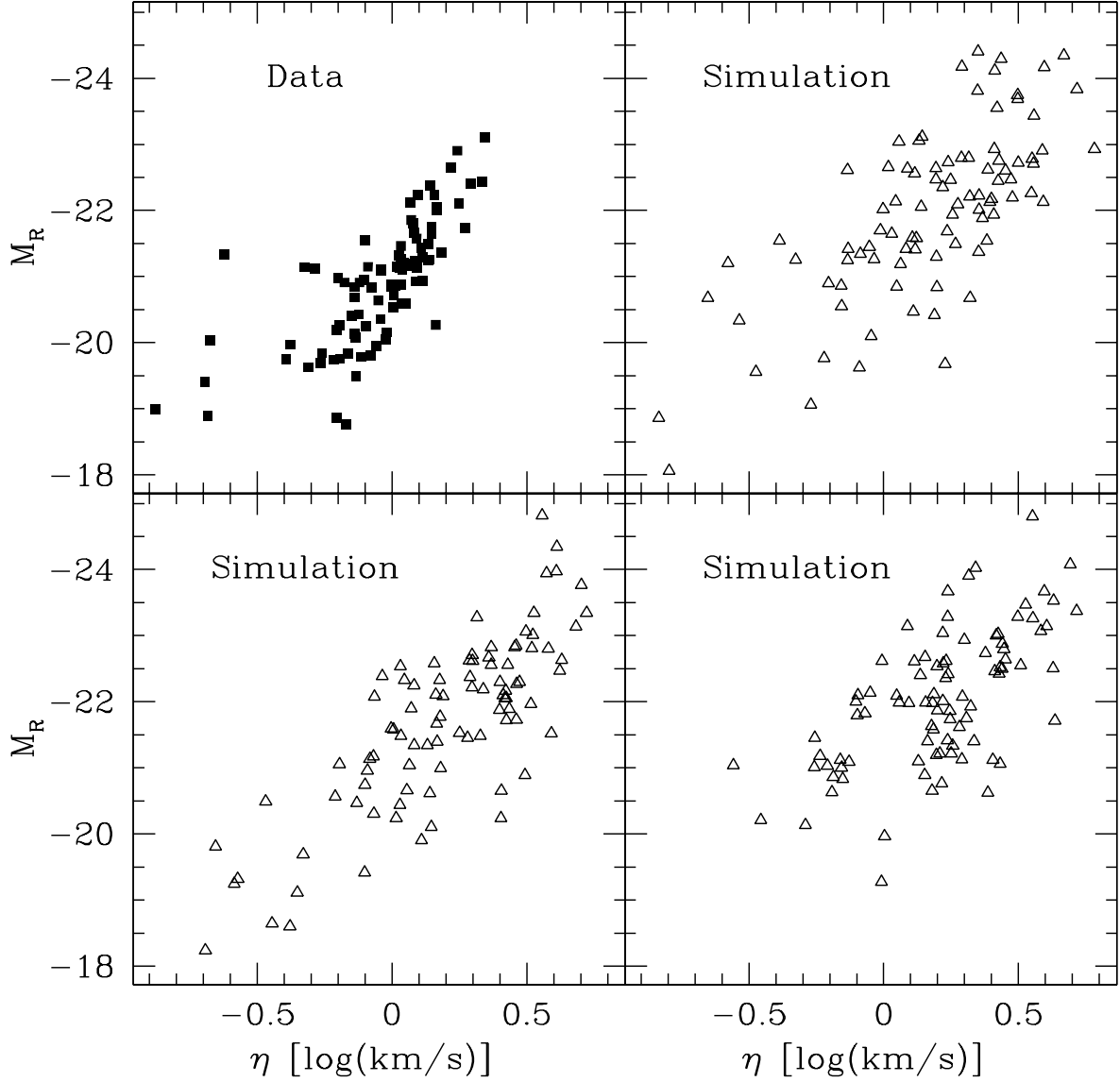}}
\caption{Comparison of the TF distribution of the pair data and three Monte
Carlo realizations: (a) the full dataset, and (b) -- (d) three realizations
using the TF parameters and scatter from the full dataset.  The distributions
appear different from the data --- the high-mass scatter in the data
is much smaller than that of the simulations.  Thus, we infer that the
low-mass scatter is due to a non-Gaussian tail of outliers.}
\label{fig:fakeout}
\end{figure}

\begin{figure}
\centerline{\epsfxsize=6.5in%
\epsffile{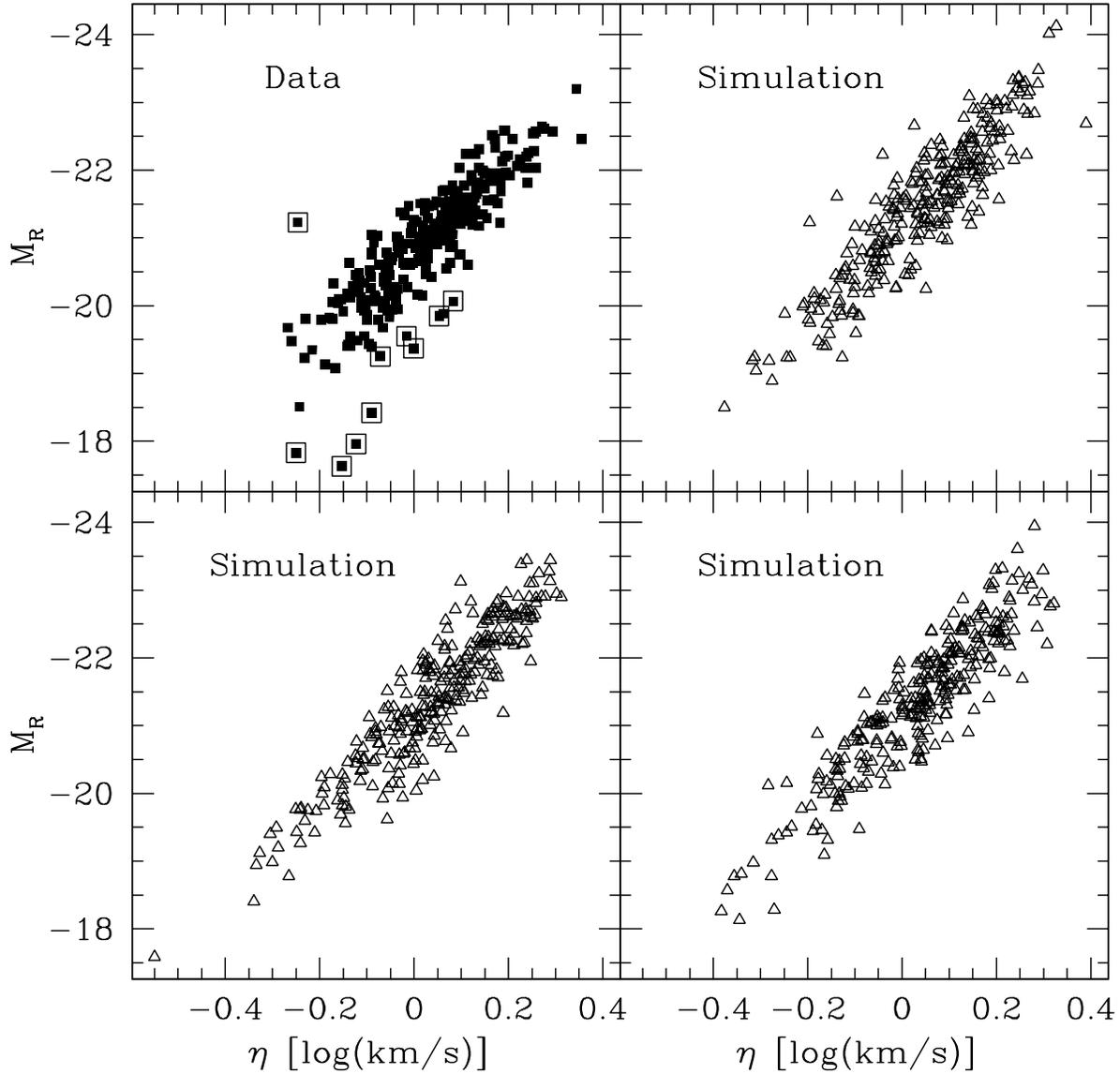}}
\caption{Comparison of the TF distribution of the C97 data and three Monte
Carlo realizations: (a) the full dataset, with 3$\sigma$ ``outliers'' marked,
and (b) -- (d) three realizations using the TF parameters and scatter 
from the data {\it with the outliers}.  The Monte Carlo realizations
are similar to the TF distribution of the data points; 
there is no evidence that
the marked points are true, non-Gaussian outliers.}
\label{fig:C97noout}
\end{figure}

Fig.~\ref{fig:fakeout} shows the TF distribution of the pair
data, along with 3 examples of Monte Carlo realizations.
The data and realizations are not a good match: even though 
we used the {\it measured} scatter to construct the
realizations, the scatter in the realizations
substantially exceeds the scatter in
the data at the high-mass end.  
Thus, the observed pair distribution includes
a small set of apparent outliers to the TF relation.
(By comparison, the Monte Carlo realizations of the
C97 data are a much better match in Fig.~\ref{fig:C97noout}.)

\begin{figure}
\centerline{\epsfxsize=6.5in%
\epsffile{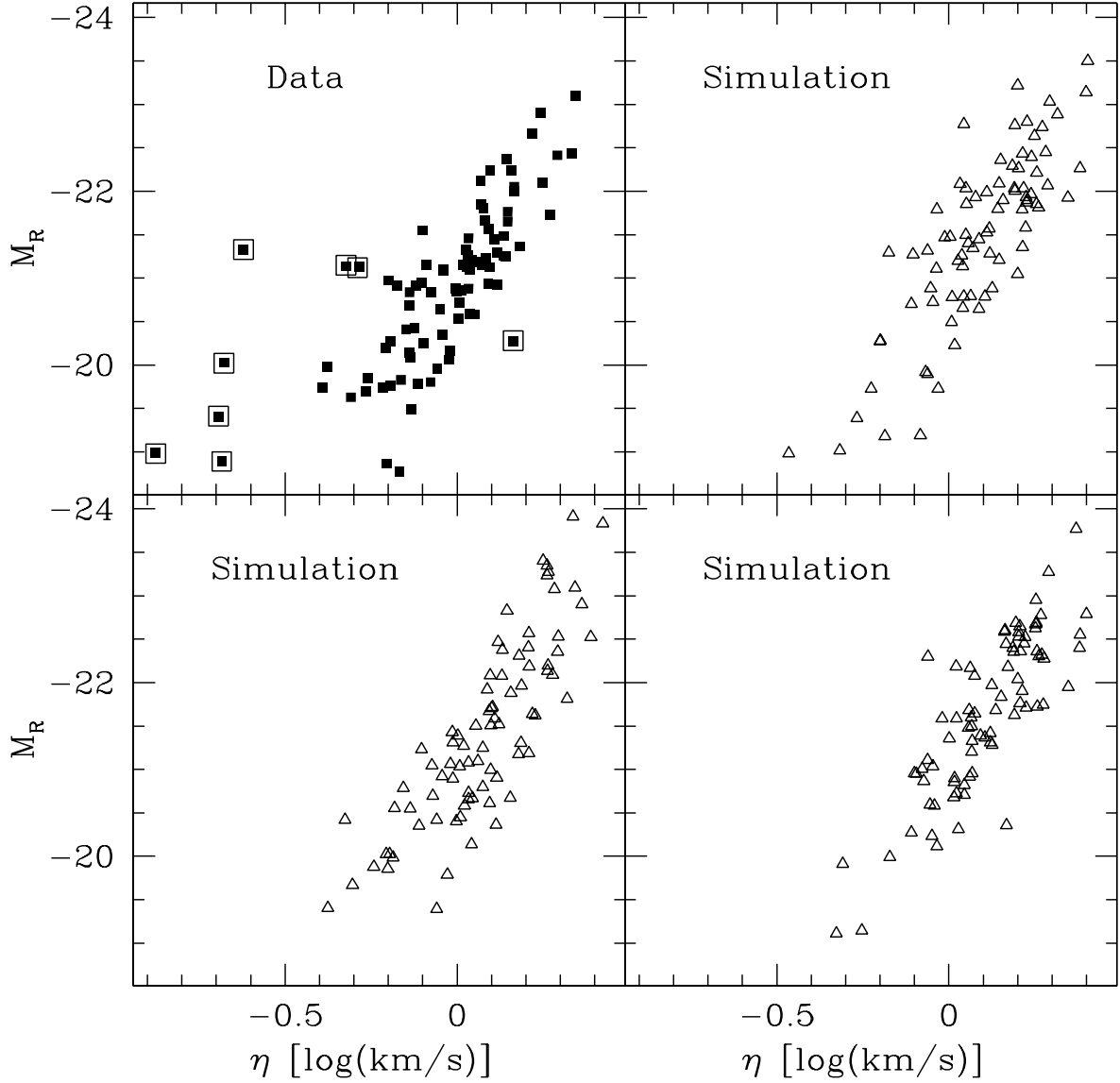}}
\caption{Comparison of the TF distribution of the pair data and three Monte
Carlo realizations: (a) the full dataset, with outliers marked,
and (b) -- (d) three realizations using the TF parameters and scatter 
from the data without the outliers.  The Monte Carlo realizations
are similar to the TF distribution of the non-outlier data points.}
\label{fig:fakenoout}
\end{figure}

We identify these outliers using a simple sigma clipping algorithm.
Because the outliers dominate the slope computation, some
assumption about the slope is necessary.  We eliminate points by fixing
the slope to the value for the C97 distribution without
bias correction, -6.36.
We use this slope for two reasons:  the slope of the observed C97 distribution 
is applicable to the observed pair distribution because the 
magnitude limits for the samples are the same, and
we seek to keep the prescription simple and easy to implement.
In practice, the choice of -6.36 instead of -7.03 does not affect
the set of outliers we identify.  We eliminate
outliers iteratively, by (1) fixing the slope and computing the
offset, $\Delta_{\rm TF}$, and dispersion, $\sigma_{\rm TF}$,
(2) removing the 3$\sigma$ outliers and the points with error
bars extending outside the 3$\sigma$ region, 
and returning to (1) until the sample is stable.  (We exclude points
with error bars outside the 3$\sigma$ region
chiefly in order to exclude one galaxy, UGC~4774, which lies
extremely close to the boundary, and is in fact a 2.94$\sigma$
outlier --- this procedure is equivalent to 2.94$\sigma$ clipping.)  
In this way, we eliminate 8 outliers to the
distribution.  Fig.~\ref{fig:fakenoout} shows that
the Monte Carlo realizations now match the 
slope and scatter of the non-outlier data very well.  
These 8/90 outliers represent a relatively high number;
for example, TP00 identify only 2/157 outliers (at a 
somewhat higher 4~--~5-$\sigma$ level).

Columns (5) -- (7) in Table~\ref{tab:tf}
and  the 3rd row in Table~\ref{tab:tfparams} show the TF 
properties of these 82 galaxies.  
We plot the TF parameters of the sample minus the outliers
in Fig.~\ref{fig:tfparams}.  
The solution for the non-outlier galaxies
is closer to the C97 distribution.  

Based on the 68.3\% confidence errors (see
the description of Table~\ref{tab:tfparams}),
the zero points of the TF relations differ very little.
The ``field'' galaxies are
0.22 magnitudes {\it brighter} (with only
1.4$\sigma$ significance).  As we show in Sec.~4,
most TF deviations expected in interactions would act
in the opposite direction: the pair galaxies would appear
``overluminous''.  Thus, the marginally
significant zero-point offset may be unrelated to
the interactions. 

The conversion from Gunn $r$ to Cousins $R$ of
0.354 magnitudes from J$\o$rgensen (1994) 
is one source of uncertainty; she reports a scatter
of 0.035 magnitudes from a set of 32 standard star
measurements, and a color term of $-0.111(g-r)$ for
the reverse measure, $R-r$ (which we do not implement).  
The color term implies that 
color differences among samples could lead to substantial
systematics introduced by the conversion.
Frei \& Gunn (1994) measure $r-B$ and
$R-B$ colors, and find that $r-R$ ranges from
$0.31$~--~0.33 for ellipticals
through irregulars using
spectral energy distribution templates.  Because of differences
in the literature and uncertainties in the color term,
errors from this conversion of up to $\sim$0.05 magnitudes
are possible.  In addition, Monte Carlo results
suggest that our failure to account
for diameter limit bias in the C97 sample could 
(artificially) decrease the measured zero points
by a further $\sim$0.06 magnitudes.

The slopes differ more significantly: 
we find -5.58 for the pairs with outliers
removed and -7.03 for the complete C97 sample; the difference is
1.45 with a nominal 2.6$\sigma$ significance.  
The actual significance of this slope difference is 
probably greater
because our slope-constrained outlier clipping forces
the slopes to agree. The
choice of fixed slope for clipping forces the pair slope to
steepen and, if applied to the C97 sample, would force
the C97 slope to become shallower.
In spite of these ambiguities,
the pair TF parameters are relatively 
stable to our choice of 3$\sigma$ clipping.
Even when we clip by 2$\sigma$, removing 14 of the galaxies,
the pair TF slope steepens by  $\sim$0.32, or $< 1 \sigma$.

Additional arguments support the difference between the C97 and
pair TF distributions: (1)  the pair distribution outliers 
are twice as frequent as the C97 ``outliers'': 9\% of the
pair galaxies are $\geq 2.9\sigma$ outliers, compared with 4\% (11/279) of the C97
galaxies [compared with an expected $\leq 5/297$ outliers 
from the Monte Carlo simulations, with 99\% confidence]. 
More important, (2) the pair outliers are apparently
``overluminous'', whereas 10/11 of the C97 ``outliers''  (which we
do not remove for our computations) lie below
the TF distribution.  Figs.~\ref{fig:C97noout} and
\ref{fig:fakenoout} thus
highlight the true differences in the TF distributions.

In summary, the C97 and pair TF distributions differ somewhat;
the outliers and the modest slope differences provide the
strongest clues to the true differences.
We discuss the implications of the slope differences
in Sec.~5.2.

The outliers have important implications
for the study of the TF properties of interacting galaxies
and for TF studies of galaxies at high redshift where the
detailed properties of the galaxies may be unknown.  
Next, we explore the physical origin 
of these outliers
which may result from luminosity evolution {\it or}
from errors in the measurement and/or interpretation of
the TF parameters.  

\section{Scatter and Offset in the TF Relation}

Several studies use the TF relation as a measure of cosmological
luminosity evolution (e.g., Vogt et al. 1996, 1997; Simard \& Pritchet 1998).
However, luminosity evolution is not the only factor affecting 
the TF properties of a galaxy.
The physical processes which cause substantial
luminosity boosts may affect the TF parameters
in other ways.
If these processes are at work during
galaxy interactions, the perturbed galaxies may depart
from the TF relation, at least temporarily, due to factors
other than or in addition to a luminosity boost (before perhaps
resettling onto the relation).

\begin{figure}
\centerline{\epsfysize=7.2in%
\epsffile{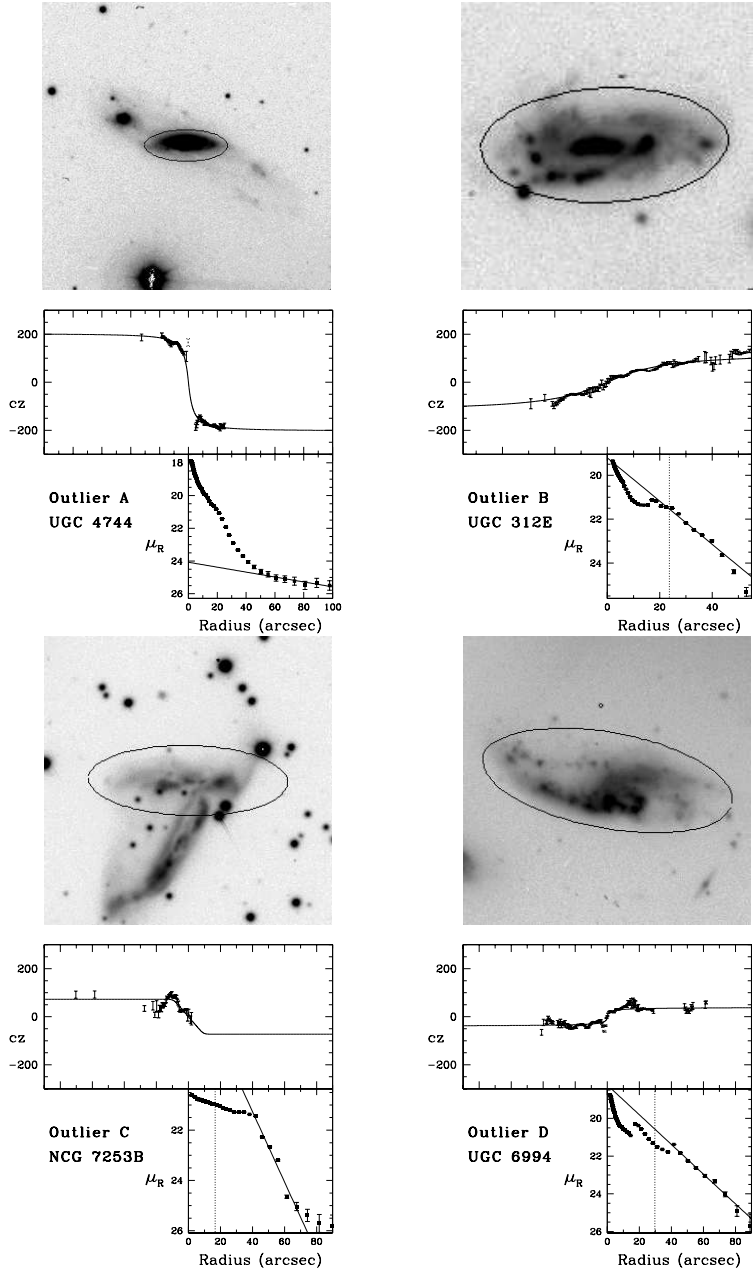}}
\caption{Outliers to the C97 TF relation.
For each galaxy,
the top panel is the $B$-band image, with the ellipse we use
to measure $\epsilon$.  The middle panel is the rotation curve,
on the same spatial scale as the image, and the bottom
panel is the $R$ surface brightness.  The vertical
line marks 2.15 R$_{\rm disk}$, the spatial position used to
measure V$_{\rm c}$.}
\label{fig:out1}
\end{figure}

In this section, we start with the most the obvious cases
of deviations from the TF relation,
the outliers (Sec.~3.2), and explore the clues to their physical origin.
We label these outliers ``A'' --- ``H''
in Fig.~\ref{fig:c97} and we list
their properties in Table~\ref{tab:outliers}. We show their
images, rotation curves, and surface brightness profiles in
Figs.~\ref{fig:out1} and \ref{fig:out2}.
In Secs.~4.1 -- 4.3, we consider the {\it potential} effects of the 
interaction on galaxy TF parameters, and discuss the outliers
in light of these effects.  We summarize in Sec.~4.4.

The effects include:  
(1) galaxy overlap, which may make separate magnitudes difficult
to measure, 
(2) peculiar velocities,
(3) tidal stretching, which 
pulls flux below the surface
brightness limit, artificially elongates
a face-on galaxy to appear more edge-on than it really is and affects
our ability to measure the photometric inclination, or
affects measurements of the disk scale length, needed to
compute the C97 V$_{2.2}$ velocity-width statistic, 
(4) kinematic distortion, which results in a distorted 
or truncated rotation curve, and 
(5) a luminosity boost from star
formation associated with the interaction.

Effects (1) and (2) introduce measurement errors which are
relatively straightforward to estimate (see Sec.~2.3) --- we
include these effects in the magnitude errors. They may
increase the general scatter in the TF distribution, but
they are not responsible for the outliers.
Effects (3) and (4) are more elusive; 
we group them into the
category of ``parameter misinterpretation errors''.
These errors occur when measurements of galaxy properties
do not represent the same
physical quantities that they would for an isolated
galaxy in equilibrium.
For example, a close galaxy-galaxy pass may
distort a galaxy so
that the measured kinematics 
reflect the transient phenomenon,
not the gravitational potential of the galaxy.
{\it The error bars do not reflect parameter misinterpretation
errors.}  We discuss the contributions from these errors below.

\subsection{Tidal Distortion}

Tidal elongation affects the TF properties of the
interacting galaxies.  Franx \& de Zeeuw (1992) describe
the way moderately elongated 
galaxies (with triaxial or prolate dark matter halos)
affect the TF relation (see also Beauvais \& Bothun 1999;
Bershady \& Anderson 2000).
However, they restrict their analysis to galaxies with
small intrinsic elongations; we require an estimate of
the effects of large elongations on photometric parameters.

To illustrate the effects from tidal distortion,
we use results from the numerical simulation
of Barton, Bromley, \& Geller (1999), who demonstrate the effects
of this transient distortion on galaxy rotation curves.
Fig.~\ref{fig:t60} shows a simulation of an interacting
galaxy from Barton et al. (1999); it is a ``Milky Way B''
model (see Kuijken \& Dubinski 1995), with an intermediate
halo size and moderate impact parameter,
after a prograde encounter with an equal-mass galaxy.  Here,
we examine the ``toy'' model, with 100,000 particles
per disk, rather than the fully self-consistent model.  We
choose this timestep because the kinematic and morphological
distortions are dramatic.  The initial model is a stable,
axi-symmetric exponential disk.

The simulation in Fig.~\ref{fig:t60} is face-on --- 
the ``true'' inclination angle is $\sim 0^{\circ}$.  However,
the tidal interaction elongates the galaxy.  
In Fig.~\ref{fig:fits} we show the results of applying
the photometry software to the initial galaxy (circles)
and the galaxy in Fig.~\ref{fig:t60} (triangles).  Open
points show ellipses fit by the program; closed
points show ellipses fixed by the program or that represent
non-convergent fits.   Although 
Fig.~\ref{fig:fits} applies to only one 
timestep in one simulation, it shows a 
relatively strong tidal distortion and should
represent an approximate upper limit to the distortion
introduced by a tidal encounter.

The tidal stretching has little effect on the
observed total (face-on) magnitude of the galaxy --- 
only the fainter parts of the outer disk depart from the 
initial galaxy.
A stronger encounter could do more damage, but it might
affect the morphology of the galaxy enough
to exclude it from a sample of disk galaxies.
Similarly, the surface brightness profile, hence
the face-on measured disk scale length, remains regular,
in part because the program allows the 
position angles and ellipticities to
vary as a function of radius.

The effects of the interaction on the intrinsic 
ellipticity of the galaxy, 
which we define as the ellipticity of the galaxy 
when viewed face on, are much more
apparent.  The ellipticity reaches
0.4885, corresponding to an inferred inclination of $60.9^{\circ}$
(compared to a true inclination of 0$^{\circ}$).  This 
elongated galaxy
would appear in our TF sample, but it would have
a velocity width near zero and would therefore be
an apparently overluminous outlier to the TF distribution.

In Fig.~\ref{fig:inc} we show the offset in the velocity
width, $\eta$, which results from tidal elongation.  We compute
$\Delta \eta$ assuming $\sim$2600
observation angles spaced uniformly about the unit
circle.  For the calculation, we assume a thin, 
elliptical disk, compute the projected figure of the ellipse,
and measure the position angle misalignment of the major axis, 
and the inclination, i$_{\rm meas}$.  Because we use the 
thin disk approximation, we compute with the simple formula
$\cos({\rm i}_{\rm meas})=\frac{\rm b}{\rm a}$, where a and
b are the maximum and minimum diameter of the projected
figure.  We require i$_{\rm meas} > 40^{\circ}$ to include
the ``galaxy''; we compute 
$\Delta \eta = \log[\sin({\rm i}_{\rm true})/\sin({\rm i}_{\rm meas})] + \log(f_{\rm align})$,
where f$_{\rm align}$ is the error introduced by the position angle
mismeasurement (see Equation~\ref{eqn:misalignment}).

Any errors in $\eta$ due directly to inclination mismeasurement
from a non-zero intrinsic ellipticity are accompanied by small
errors in absolute magnitude due to an incorrect
inferred reddening correction, and larger errors in the
measured disk scale length, which Fig.~\ref{fig:major}
describes for idealized ellipses.  These have more
minor (second order) effects on the measured  TF parameters.

Tidal elongation introduces appreciable errors only in
extreme cases.
The second panel in Fig.~\ref{fig:inc} is
appropriate for the simulated galaxy in Fig.~\ref{fig:t60}.
Even for this extreme example, 
errors in $\eta$ would exceed 0.3 dex only 24\% of
the time.  Furthermore, 
these distortions are short-lived ($\lesssim$ a few hundred Myr).
Because we sample pairs both
before {\it and} after a close pass, at most $\sim$half 
of the sample could be post-close-pass.
In addition, these features
are only associated with strong, prograde interactions, which will
occur only a small fraction of the time ($\ll$ half).

We conclude that only a small number of outliers ($\lesssim 3\%$) 
would result from the elongation effects of tidal forces.
They should have tidal-tail morphology.  
Of the outliers in Figs.~\ref{fig:out1} and \ref{fig:out2},
only UGC~6994 has the distinctly tidal ``grand-design'' structure;
it is therefore the best candidate for inclination 
mismeasurement due to tidal
distortion.  Because tidal distortion is unlikely
to affect $\eta$, the other galaxies are probably outliers
for other reasons.

\subsection{Kinematic Distortion, Truncation, and 
Narrow Emission Line Galaxies}

Fig.~1 of Barton et al. (1999) shows the transient kinematic 
signatures of tidal distortion produced in simulations of
dissipationless interactions.  The figure generally 
includes the most extreme timestep in each simulation, and
therefore represents an approximate upper limit to the
distortion in the models.
We estimate V$_{2.2}$ from the
simulations using the original disk scale length (one length
unit in the simulations). Because
the C97 rotation curve model provides a very poor fit to these
curves, we measure the velocity width directly (i.e., without
use of the model), by interpolating points in the vicinity
of 2.15~R$_{\rm disk}$.  Most of the departures from the
undisturbed model (V$_{\rm 2.2} \approx 1.8$) result from
the dips in the rotation curves. The deviations in the TF parameter,
$\eta$, resulting from these distortions range from -0.11 to 0.07 dex
for these nine simulations, with a median of -0.07 dex. 
If these relatively centrally concentrated
models represent typical galaxies, 
the expected deviation from the TF relation due
to dissipationless effects on well-sampled rotation
curves is very small.

Gasdynamical effects on the measured kinematics
are much more uncertain.
Close passes between galaxies initiate gas infall (e.g., Mihos \&
Hernquist 1996) which BGK detect as central star formation.  
The effects of the infall on H$\alpha$ kinematics 
are difficult to estimate due to uncertainties
in the hydrodynamical prescriptions adopted in numerical simulations.
A comparison between stellar and gas kinematics in
individual galaxies would aid in estimating the magnitude
of this effect.  

Non-uniformity in the spatial distribution of the ionized
gas is probably the most substantial cause of errors in
linewidth interpretation.  
If the gas is concentrated in the center
of a galaxy (as expected after a close pass), or in 
lumpy emission on only one side of the galaxy, 
the rotation curve will not be well sampled.  
These effects generally result in an underestimate
of the velocity width.

We can estimate the importance of non-uniform 
H$\alpha$ emission in individual cases. NGC~7253B is
an obvious example --- only a fraction of the rotation
curve is sampled, presumably because the slit fell on
only one (extended) emission line region.
The rotation
curves of the outliers in Fig.~\ref{fig:out2} all appear 
truncated; this truncation may result in underestimated
velocity widths (e.g., Verheijen 1997).
Four our outliers, the truncation results from 
the concentration of emission line gas in the centers
of the galaxies.   Next, we describe the properties of
these outliers in detail.

\subsubsection{The Low-Mass Outliers --- Counterparts to 
Narrow Emission Line Systems at High Redshift?}

\begin{figure}
\centerline{\epsfysize=7.8in%
\epsffile{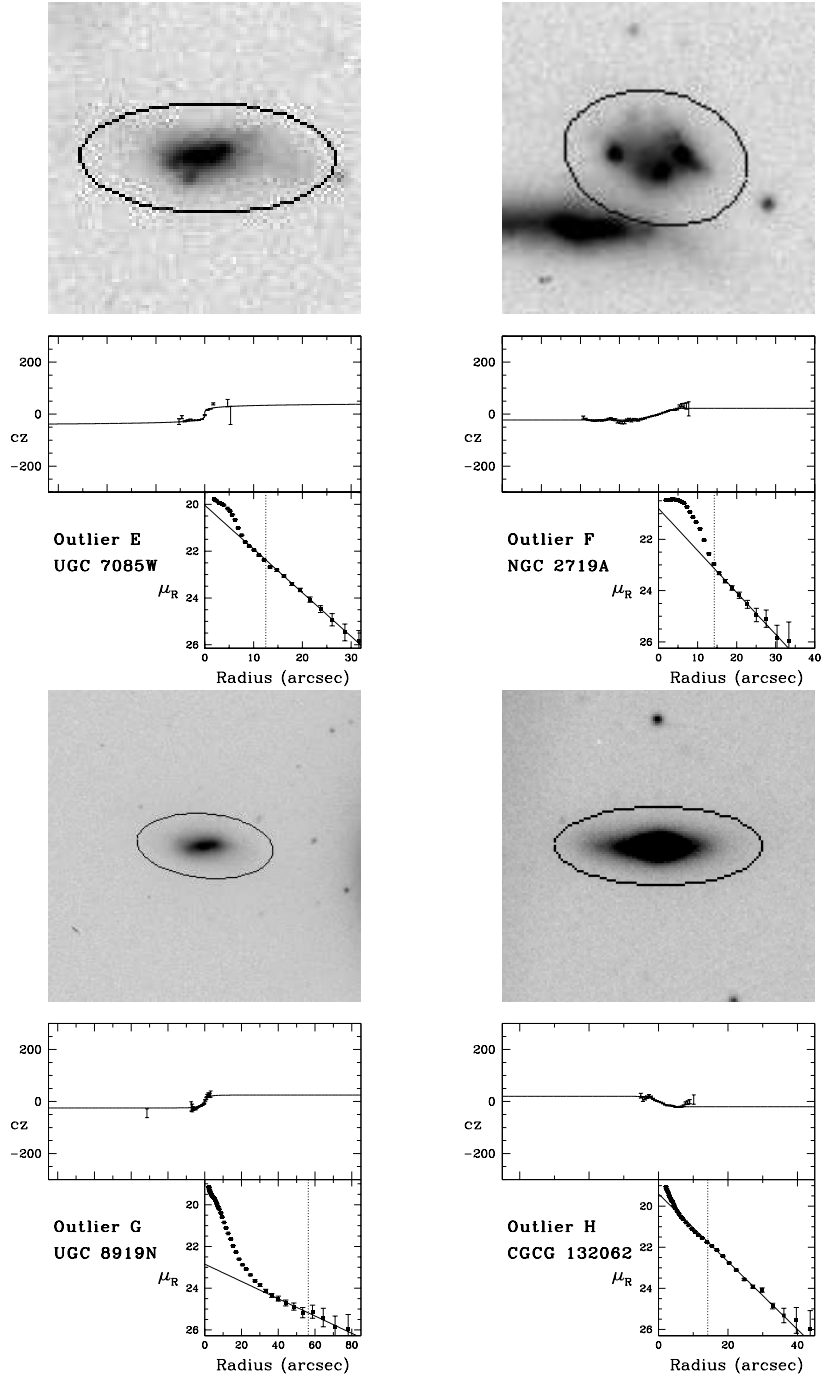}}
\caption{The low-mass outliers, ``E'' -- ``H'' 
(see Fig.~\ref{fig:out1}).}
\label{fig:out2}
\end{figure}

\begin{figure}
\centerline{\epsfxsize=6.5in%
\epsffile{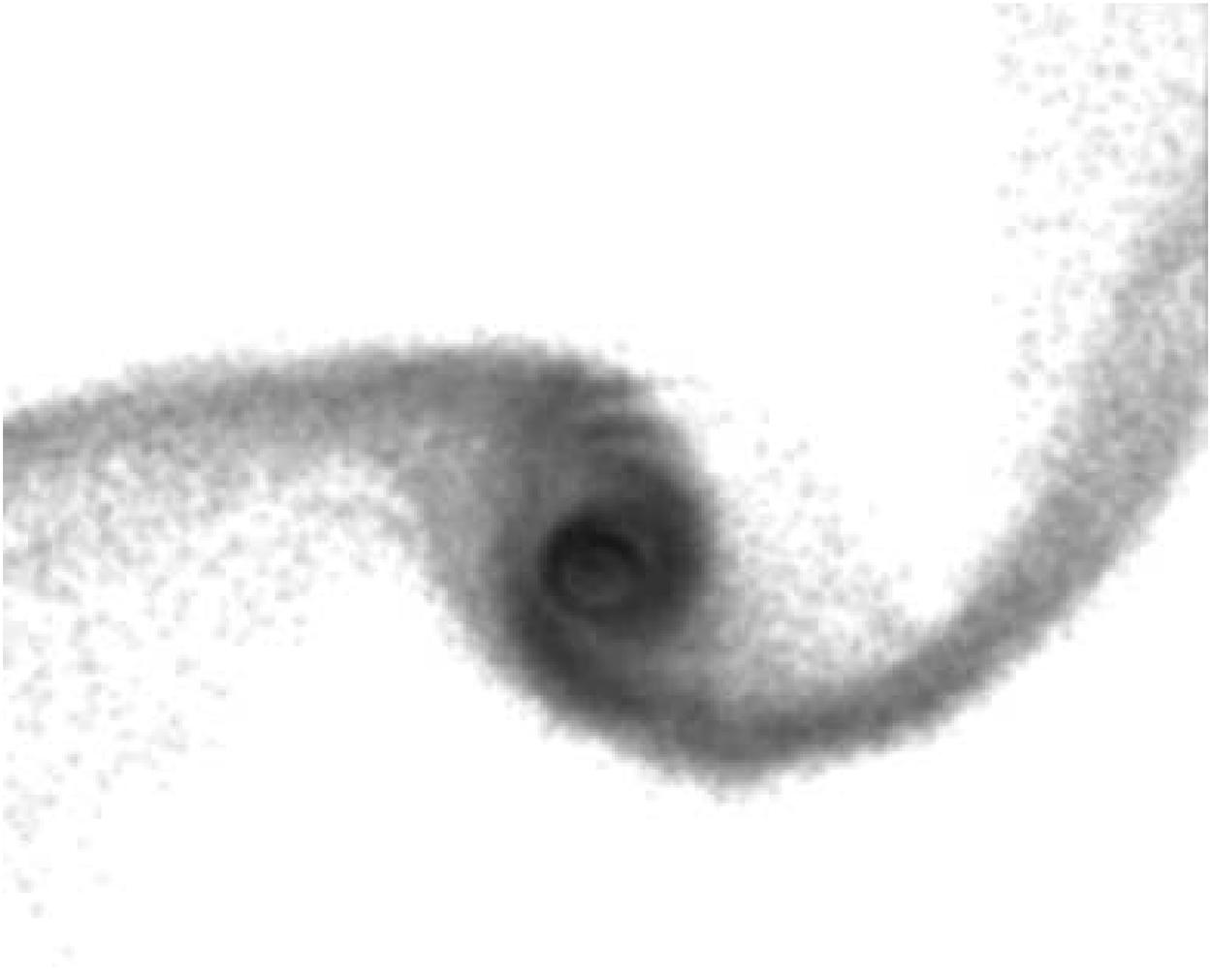}}
\caption{Simulation from Barton et al. 1999.  We show a face-on
``Milky Way B'' model galaxy with an intermediate
halo size and moderate impact parameter (see Barton et al. 1999),
after a prograde encounter with an equal-mass galaxy.}
\label{fig:t60}
\end{figure}

\begin{figure}
\centerline{\epsfxsize=6.5in%
\epsffile{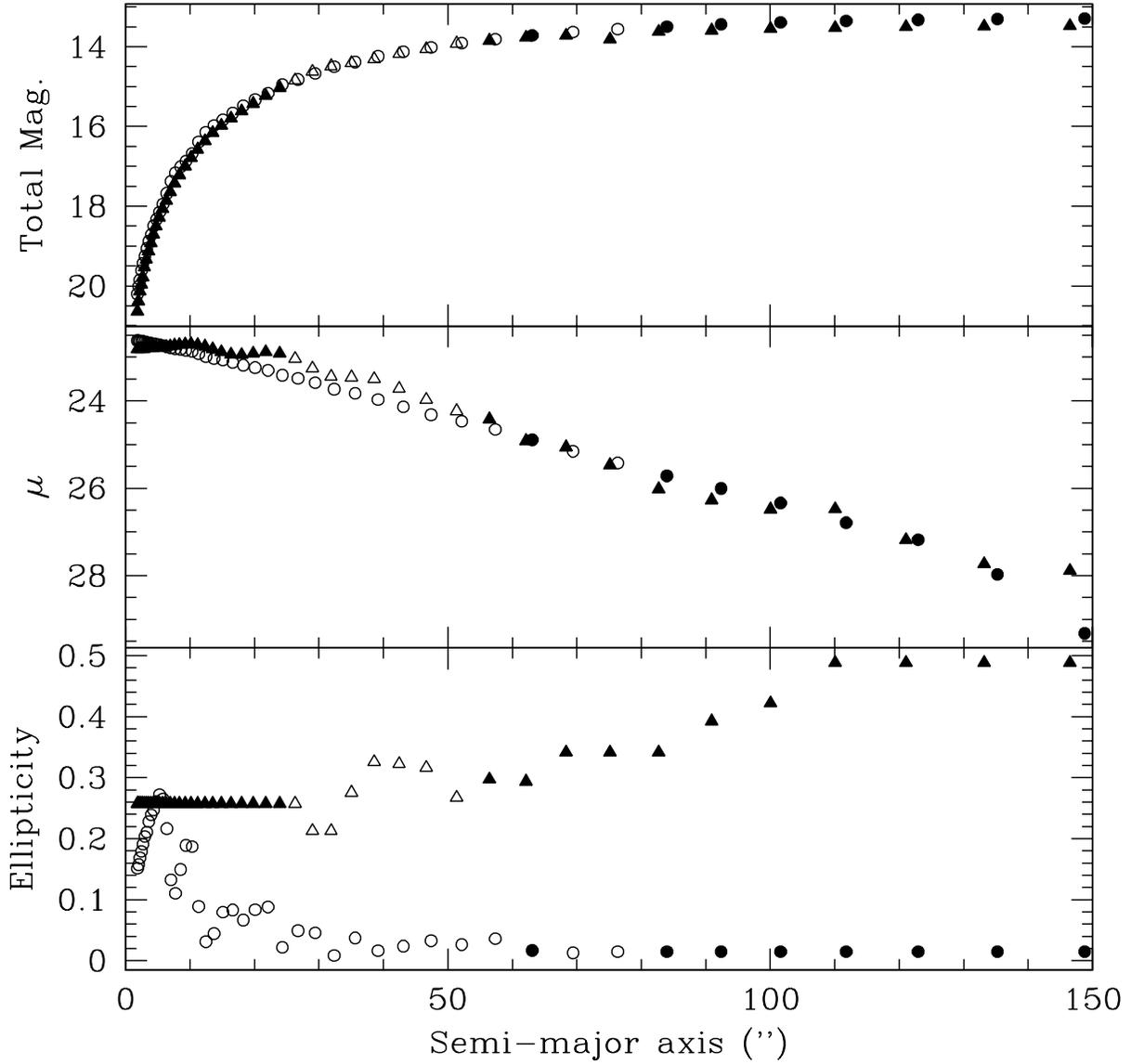}}
\caption{Results of the ``photometry'' on the simulated 
interaction (triangles) and
the initial simulated galaxy (circles).   Open
points represent ellipses fit by the program; closed
points were fixed by the program or represent non-convergent
fits. The interaction has very little effect on the total magnitude
and the surface brightness profile, but the ellipticity deviates
from the ``true'' value of 0.}
\label{fig:fits}
\end{figure}

\begin{figure}
\centerline{\epsfxsize=6.5in%
\epsffile{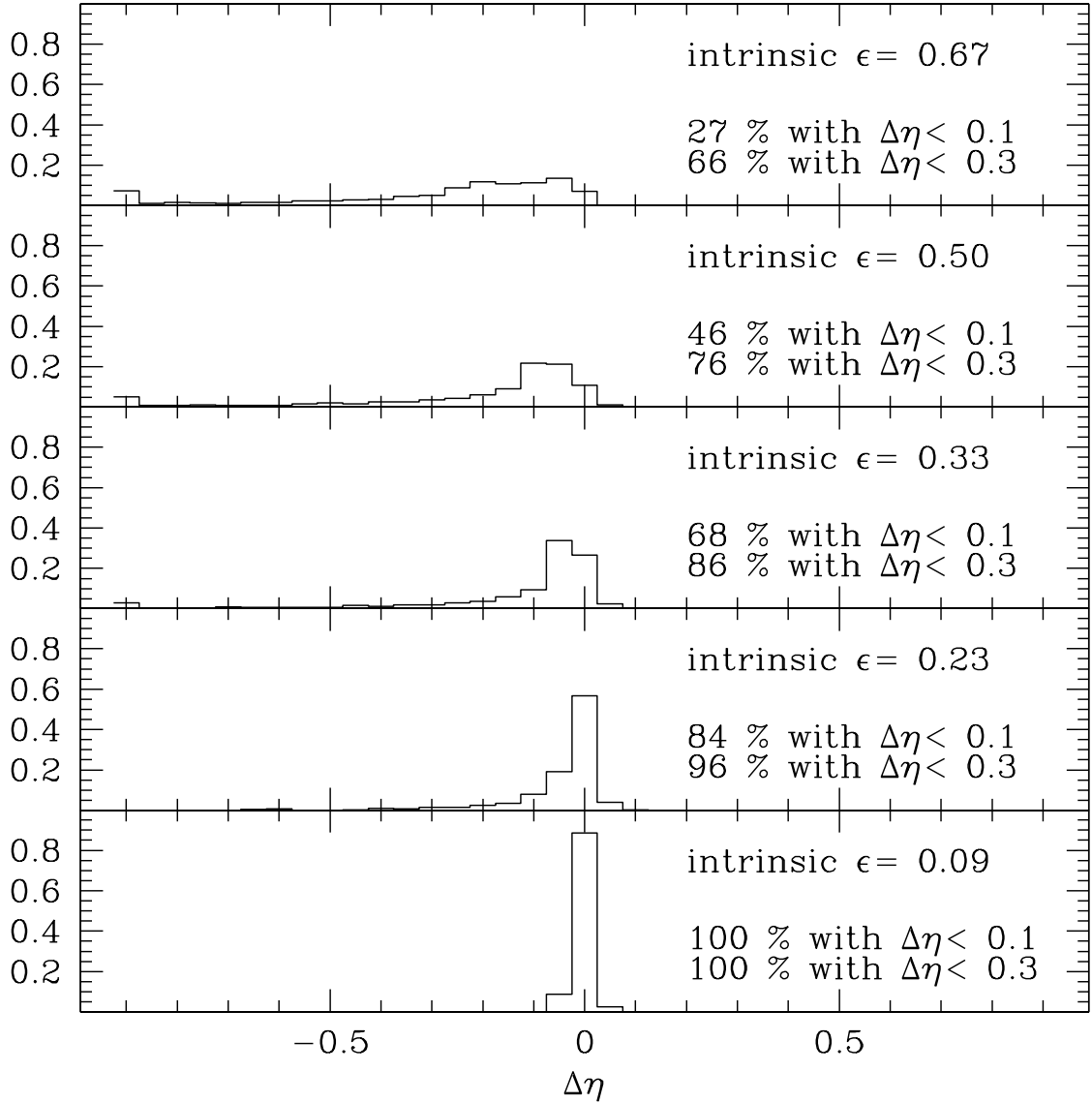}}
\caption{Effects of intrinsic elongation on the
inferred velocity width parameter, for different values of
the intrinsic ellipticity, $\epsilon$.  The histograms represent
measurements from observation angles spaced uniformly about
the unit circle; we include measurements only when 
i$_{\rm meas} > 40^{\circ}$.}
\label{fig:inc}
\end{figure}

\begin{figure}
\centerline{\epsfxsize=6.5in%
\epsffile{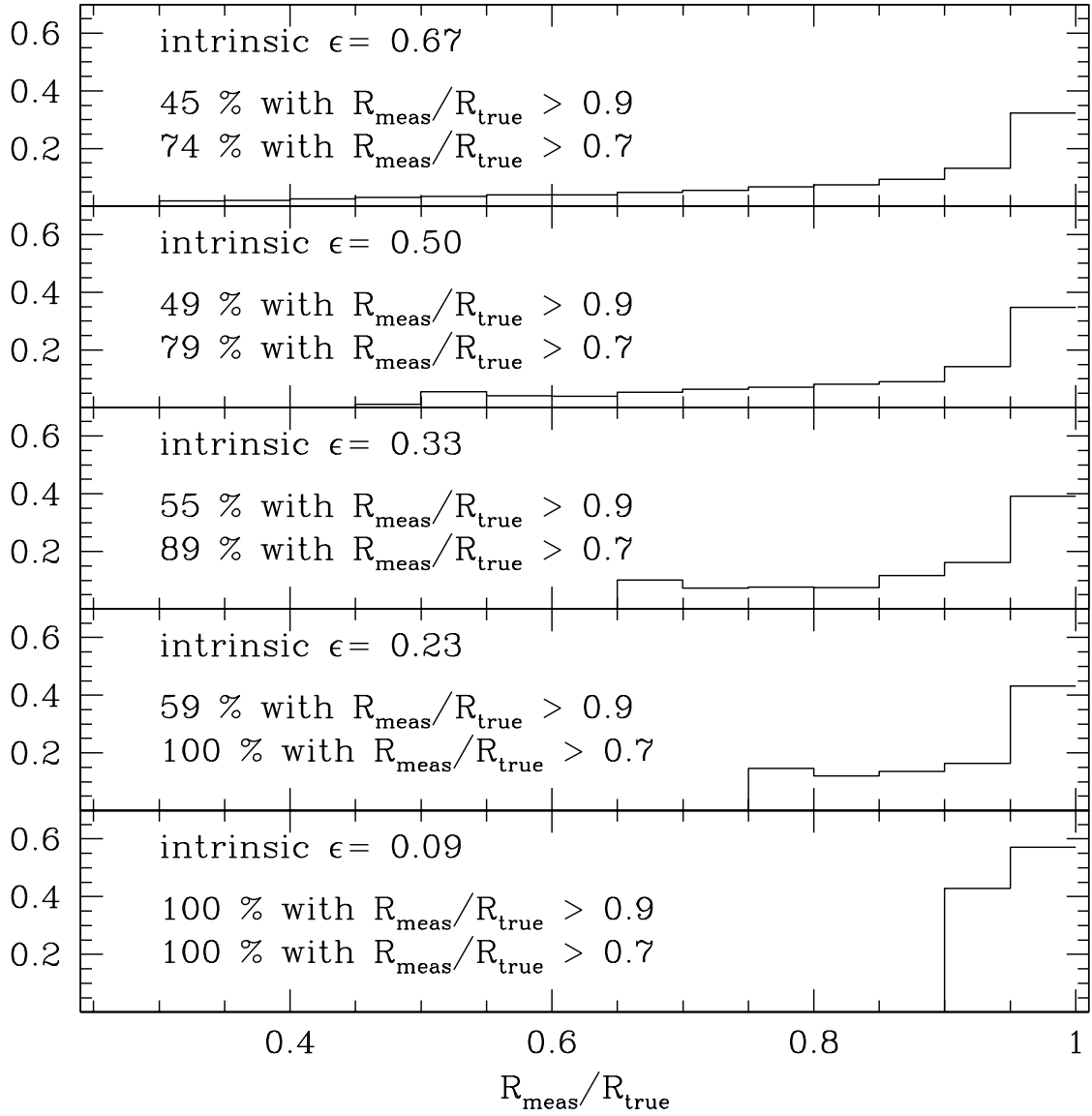}}
\caption{Effects of intrinsic 
elongation on the measured major axis length. R$_{\rm meas}$. 
As in Fig.~\ref{fig:inc}, we ``observe'' ellipses from points spaced
uniformly around the unit circle, and restrict the sample to measured 
inclinations i$_{\rm meas} < 40^{\circ}$.}
\label{fig:major}
\end{figure}

The four low-mass outliers in our study
share several properties.
Table~\ref{tab:outliers} and Fig.~\ref{fig:out2}
illustrate that these four low-luminosity TF outliers
are compact and relatively blue, with either substantial ongoing star 
formation or Balmer absorption.  
The $B-R$ colors dip in or near the centers of the galaxies, 
suggesting that the star formation is centrally concentrated, 
as expected based on simulations
of interacting galaxies (e.g., Mihos \& Hernquist 1996).

Fig.~\ref{fig:out2} shows that the rotation curves do not 
flatten substantially.  
Spatially extended, two-dimensional velocity mapping of the 
galaxies would determine whether they lie
on the TF relation.  Only UGC~8919N has existing spatially-resolved
HI observations. 

UGC~7085W has a rotation curve that extends only a few arcseconds.
It is irregularly shaped, and its colors and EW(H$\alpha$) show
evidence for recent star formation.

NGC~2719A consists of a faint, 
irregular disk of emission with three bright HII regions.
The rotation curve only traces the disk well on one side; the
emission is dominated by the bright knots.

UGC~8919N has little H$\alpha$ emission and 
no obvious signs 
of tidal distortion. As the emission 
extends only $\sim5 - 10^{\prime\prime}$ from the center, the rotation curve
may only include a bar.  Archival C and D configuration VLA observations 
(P.I.: J. Chengalur) allow us to separate the HI flux of this
minor companion to UGC~8919.  Although the flux is spatially
unresolved, the profile width (at 20\% of maximum) 
is a factor of $\sim$2 wider than our measurement of the
rotation curve
indicating that in this case, 
{\it UGC~8919N lies on the radio TF relation},
and appears as an outlier in our plot only because
the truncated optical rotation curve does not 
reflect the mass of the galaxy.

CGCG~132-062, has a distinct S0 morphology with a very 
blue bulge ($B-R \sim 0.6$, uncorrected for extinction).  
It has a very small half-light radius (0.95 h$^{-1}$~kpc).
Again, the rotation curve is not
spatially well-extended (only 8$^{\prime\prime}$~--~15.3$^{\prime\prime}$) 
and may fail to describe the velocity width of the disk.
If parameter errors are not responsible for boosting NGC~2719A and 
CGCG~132-062 off the TF relation, as they are in the case of UGC~8919,
a combination of star formation
and/or substantial morphological evolution could be responsible.  

The compact, blue galaxies at intermediate redshift
(Phillips et al. 1997), including
the compact narrow emission line galaxies
(Koo et al. 1994; Koo et al. 1995;
Guzm\'{a}n et al. 1997; Guzm\'{a}n et al. 1998; CNELGs hereafter),
may be similar to these low-mass outliers.
The CNELGs are barely-resolved, heavily star-forming
galaxies found at redshifts $0.1 \leq z \leq 1.6$; CNELGs
are very compact, with half-light radii from $\sim$1-4~kpc (see
Koo et al. 1994; 1995).
Their status is currently debated --- they may be
the spiral galaxy bulges in formation (Kobulnicky \& Zaritsky 1999),
or a bursting population that will fade to present-day dwarf
galaxies.  To date, the narrow line widths of these galaxies
(28 -- 157 km s$^{-1}$, with many $< 65$ km s$^{-1}$) provide the
strongest evidence in favor of the dwarf hypothesis.  

However, at least one of our 4 low-mass outliers represents a galaxy with
a resolved emission line rotation width that does not 
represent its kinematic mass
(outlier ``C'', UGC~8919N, with archival VLA observations; 
Barton \& van Zee 2000).  Because the
properties of our outliers are strikingly similar to those of
the lower luminosity compact objects in the Hubble deep field
(Phillips et al. 1997) and the CNELGs, they may represent systems 
in similar physical states.
In particular, both the nearby TF outliers and the
CNELGs may represent cases in which
a close pass or other non-axisymmetric perturbations caused
gas infall and subsequent star formation.  The narrow line
widths and compact morphologies at intermediate redshift
would result from strong star formation confined to only 
the inner few kpc of the galaxies, during an epoch of 
bulge enhancement.

\subsection{Luminosity Evolution}

\begin{figure}
\centerline{\epsfysize=7.2in%
\epsffile{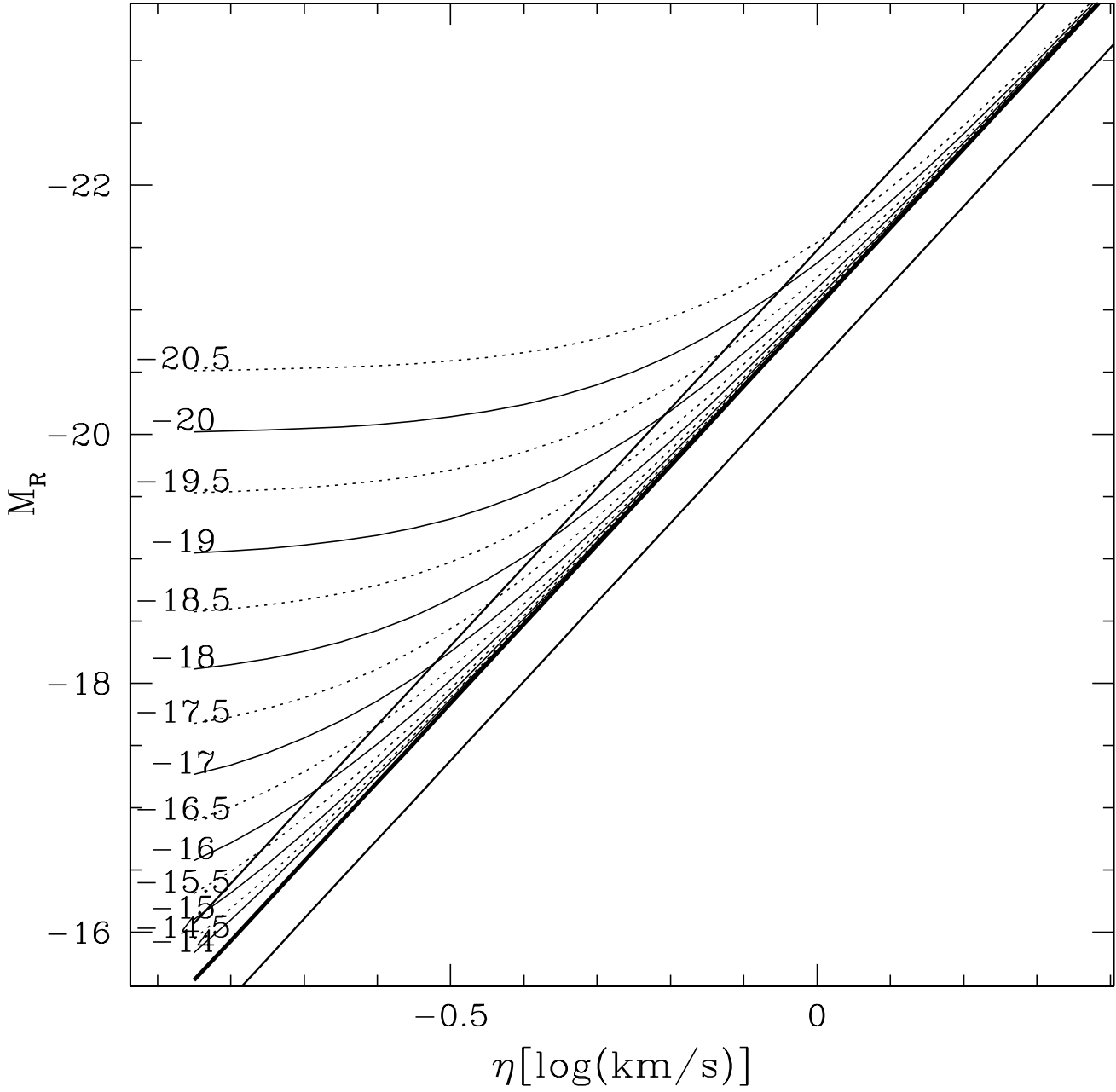}}
\caption{The luminosity effects of added starbursts: we plot 
the shift the TF properties from 
a burst of absolute magnitude M, assuming the only effect of the
burst is luminosity evolution.  
We label the contours with M, in magnitudes.  
Because the TF relation measures fractional deviations, 
a burst of constant strength changes the slope of the TF
relation.}
\label{fig:c97bursts}
\end{figure}

In Fig.~\ref{fig:c97bursts} we show the 
effects of added flux from star formation in the 
$R$ band.  We plot the C97 $R$ relation; 
the contours show the shift produced by
a burst of absolute magnitude M in the TF properties
of the galaxies, {\it assuming the only effect of the
burst is luminosity evolution}.  The figure highlights
two very important facets of the TF relation as a measure of
luminosity evolution.  First, 
the relation is steep and has a large scatter; hence,
only a large burst results in an
appreciable residual.  
Second, the Tully-Fisher relation is a relation in log space.
It measures fractional changes in luminosity and velocity width.
As Fig.~\ref{fig:c97bursts} illustrates, the 
high-mass end of the TF relation is insensitive to all but
the very brightest bursts of star formation, which are unlikely 
given the gas content of nearby bright galaxies.  Therefore,
we expect local bursts of star formation to result in {\it slope} changes.
Furthermore, it is not {\it a priori} obvious whether evolution at 
high redshift would increase the luminosity of
galaxies at a constant fractional rate:
Simard \& Pritchet (1998) find some evidence for TF
slope changes at intermediate redshift
while Vogt et al. (1997) do not.

Barton et al. (2000a) detail the evidence
for interaction-triggered star
formation in the centers of many of the pair galaxies.  
Here, we estimate the magnitude of the luminosity boost we
expect from central star formation in our sample.
The existing data only provide an estimate
of the total amount of recent ($\leq 1$~Gyr)
star formation {\it in the galaxy centers}.
Barton et al. (2000b) describe these estimates in
detail; here, we summarize the results.

\begin{figure}
\centerline{\epsfysize=7.2in%
\epsffile{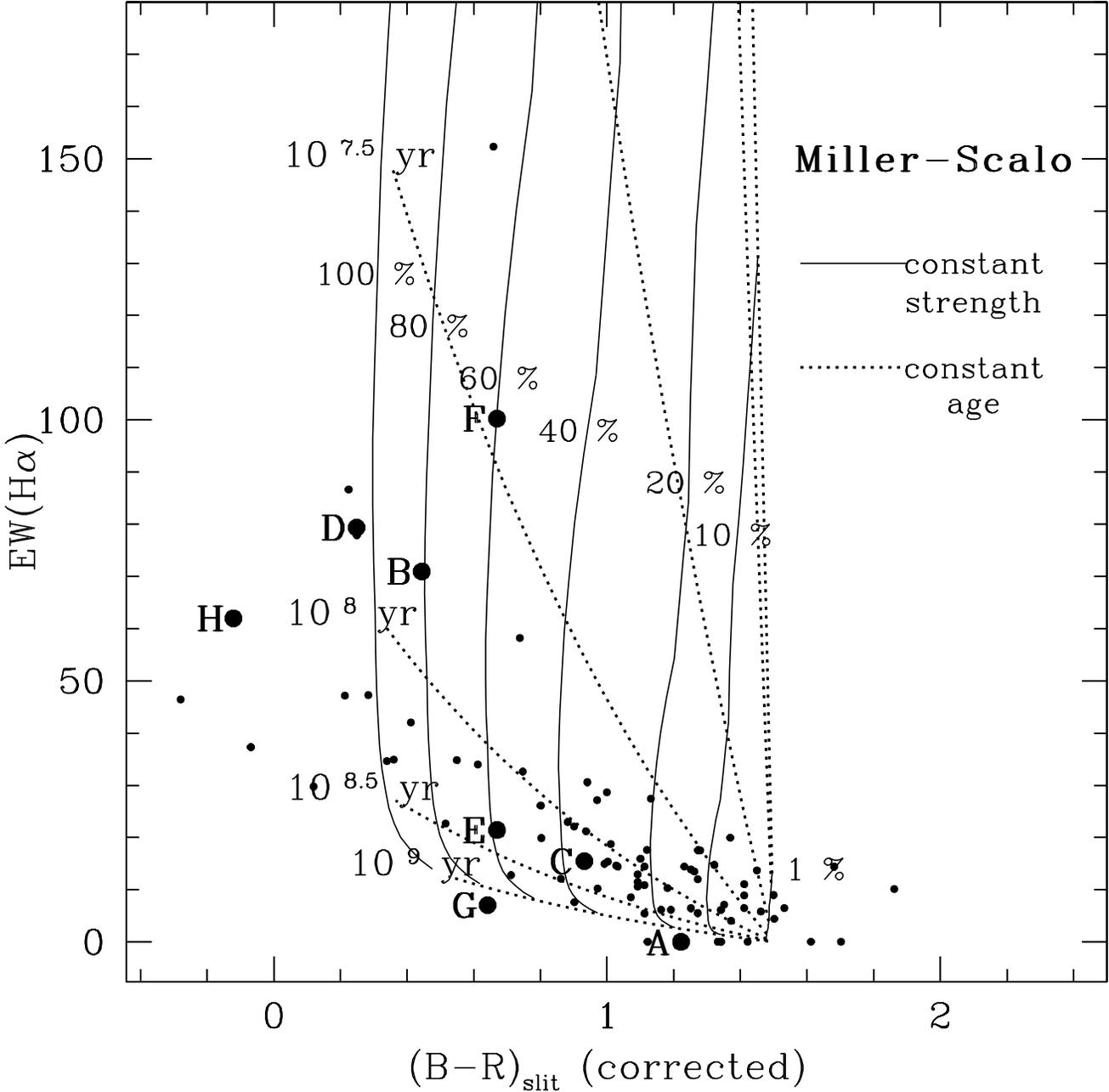}}
\caption{Central bursts of star formation in the centers of the pair
galaxies, with the TF outliers labeled.  
We plot $B-R$ colors in the central (spectroscopic) aperture vs. 
H$\alpha$ equivalent widths.  We correct the colors for reddening based on
the Balmer decrement.  The contours are lines of constant $R$-band
burst strength (solid) and age (dotted), computed with the Leitherer et al. 
(1999) models, assuming a constant star formation rate over time,
a Miller-Scalo IMF with solar metallicity, and 
$B-R = 1.5$ for the population present before 
the interaction (see Barton et al. 2000b).}
\label{fig:deburst}
\end{figure}

Fig.~\ref{fig:deburst} shows the parameters of the
central bursts of star formation
for 89 of the pair galaxies (one is absent from the
BGK spectroscopic sample).  
We use photometry in the spectroscopic
apertures (typically only $\lesssim$1~h$^{-1}$~kpc~$\times$ a 
few h$^{-1}$~kpc) to compute  $B-R$ colors, which we 
correct for reddening based on the Balmer decrement.  
The contours are lines of constant $R$-band
burst strength and age.  We use the Starburst99 models (Leitherer et al. 
1999) to compute these parameters.  We assume that all the H$\alpha$
flux comes from a recent, new burst of star formation, which proceeds
at a constant star formation rate with a
Miller-Scalo initial mass function (IMF) and solar metallicity.  We assume
$B-R = 1.5$ for the central pre-burst population 
(see Barton et al. 2000b).
Assuming that the galaxy was 
on the C97 TF relation before the star formation began,
we list the central burst strengths (fraction of $R$-band flux),
and luminosities in Table~\ref{tab:origins}, along with the ``expected''
TF residual from this central burst.  We compute this expected
residual assuming that the galaxy lay on the C97 TF 
relation before being
boosted by star formation with luminosity M$_{\rm R,burst,slit}$.

For all of the outliers, the expected boost from the central 
star formation we measure is substantially smaller than the measured residual 
(column 6).  Although the spectroscopic aperture misses much of
the flux of the galaxy (see column 4), we expect frequent centrally
concentrated star formation (e.g., Mihos \& Hernquist 
1996).  This central star formation it is not primarily 
responsible for the set of outliers we detect, although it may be
responsible for the discrepancies between the C97 ``field''
TF relation and the relation for the non-outlier pair galaxies.

\subsection{Summary of the Outliers}

In sections 4.2 and 4.3, we suggest physical origins for
the TF residuals of 6 of the 8 outliers.  
With the exception of the elongated UGC~6944, and possibly UGC~312E,
our arguments imply that neither morphological
distortions nor dissipationless kinematic distortions have
strong effects on the TF properties of these galaxies.
Nor does central star formation contribute enough flux to
explain the outliers.  
The remaining explanations for the outliers
are dissipative effects on the rotation curve 
(gas infall and non-uniform or truncated emission)
and disk star formation.  
In 5 cases, rotation curve distortion leads us to suspect
dissipative effects on the rotation curve.
Four of these galaxies are 
low-mass outliers with truncated rotation curves
that share several properties.
We identify this class of objects with 
the compact, blue galaxies at high redshift
(Sec.~4.2.1).  Finally, two of the outliers are ambiguous;
we discuss these galaxies below.
We summarize these conclusions in Column 7 of Table~\ref{tab:origins}.

UGC~4774 is the only apparently underluminous outlier,
and as such, it is the most difficult to explain.
Parameter errors
could only result from: (1) underestimating the 
inclination, which would move the galaxy at most
0.034 dex in the -x direction on Fig.~\ref{fig:c97}
due to velocity projection
and 0.3 magnitudes in the +y direction due to an underestimate
of the internal extinction (based on the C97 prescription),
(2) over-estimating the velocity width, which is
unlikely given the well-sampled rotation curve, 
(3) missing significant flux.
Its scale length, 8.1 h$^{-1}$~kpc,
is among the largest in the sample, and reflects the tidal
stretching --- all four galaxies 
with scale lengths $ > 8$~h$^{-1}$ kpc have distinct tidal
tails.
However, Sec.~4.2 demonstrates the difficulty of missing
significant flux ($\geq 40\%$) from this effect.  The final
possibility is
(4) under-estimating the Hubble-flow velocity (i.e., if its
true recession velocity is closer to 3000~km~s$^{-1}$ than the measured
2330~km~s$^{-1}$).  UGC~4774 is near the close edge of a void, 
and may be flowing away from the void and towards the Milky Way,
but its TF residual requires an unlikely peculiar velocity of at 
least 600~km~s$^{-1}$ to shift it half a magnitude.
As each single possibility is unlikely to result in a large shift,
the cause for its TF residual is either a combination
of these factors or an unknown, and more fundamental
physical difference from the other galaxies.  

UGC~312E is also ambiguous.  Although
it shows no evidence for tidal tails, it may still be
tidally distorted.  Its residual may also result
from several effects added together, possibly
including an underestimate of the scale length and 
inclination, and/or star formation in the outskirts
of the disk (not included in the spectroscopic aperture,
hence missing from Table~\ref{tab:origins}).

\section{Exploring the Origin of the Slope Offset: the Non-Outlier
Data}

In Sec.~4, we explore the expected effects of tidal distortion
and star formation on the TF relation for galaxies in
pairs.  Here, we apply empirical tests for
correlations between the TF residuals
and various measures of interaction-triggered distortion in
the non-outlier data.   

\begin{figure}
\centerline{\epsfysize=7.2in%
\epsffile{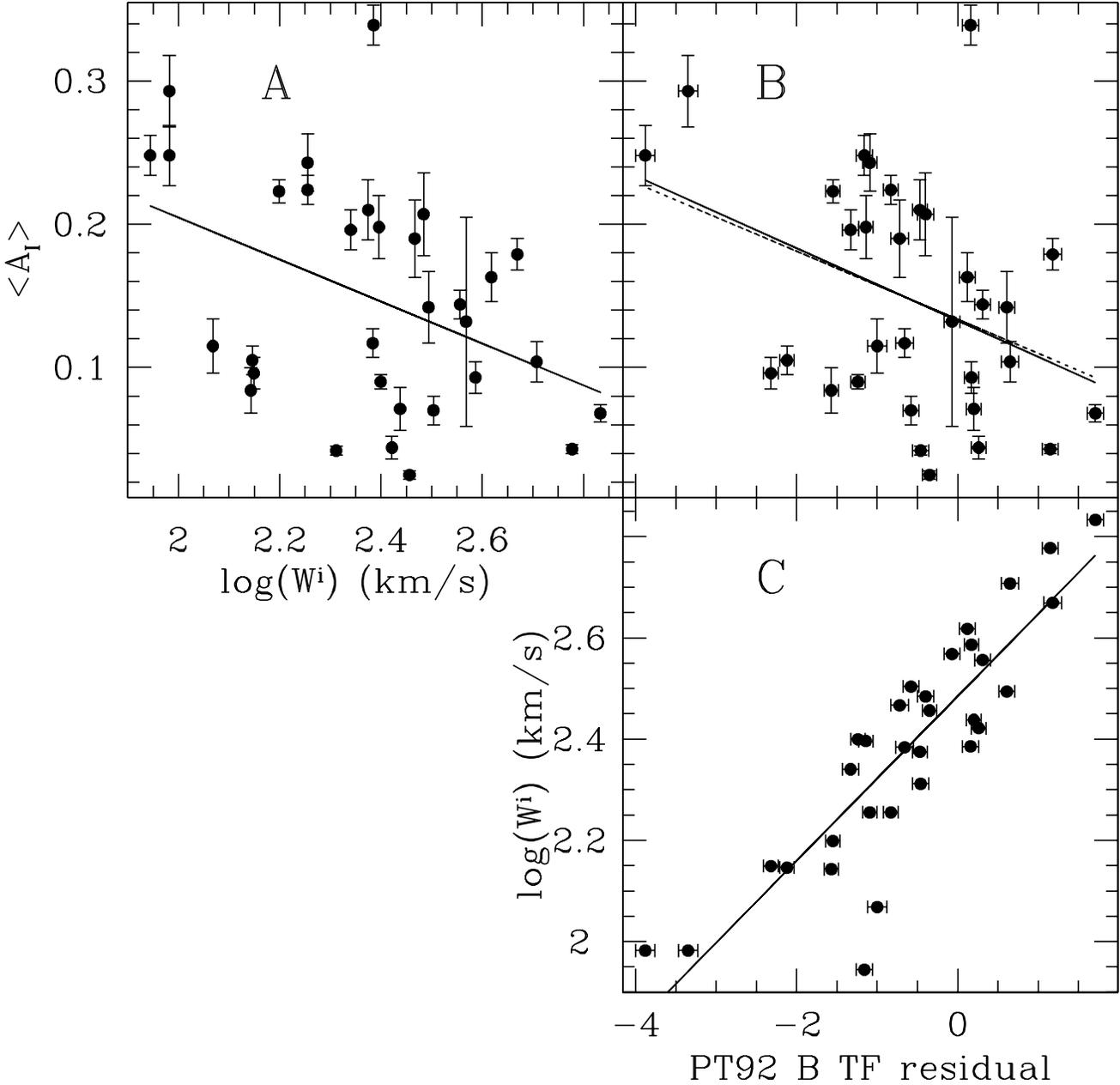}}
\caption{Systematic effects on TF residuals in the Zaritsky \&
Rix (1997) data.  The figure shows that the correlation in
Panel B (between TF residual and galaxy asymmetry)
results from the more fundamental correlation between
linewidth, W$^{\rm i}$, and asymmetry, in panel A, and the
systematic error in the Pierce \& Tully (1992) TF relation,
shown as a tight correlation between the residual and 
W$^{\rm i}$ in panel C.}
\label{fig:zr}
\end{figure}

If the third parameter 
depends on luminosity and velocity width,
identifying correlations between TF residuals and
third parameters is difficult: both systematic
errors in the slope of the relation and selection
effects can easily give
rise to false correlations.
For example, Zaritsky \& Rix (1997) report a 
correlation between photometric asymmetry and 
the residual from the Pierce \& Tully (1992) $B$-band
TF relation (panel B of Fig.~\ref{fig:zr}); the correlation might
result from increased star formation in
asymmetric galaxies (see Rudnick, Rix, \& Kennicutt 2000).
If real, this conclusion would
be discrepant with our results (Sec.~5.1). However,
the correlation is an artifact.
Panel C of Fig.~\ref{fig:zr}
shows the relationship between the $B$-band TF residual
from the Pierce \& Tully (1992) relation
and the velocity width. The figure shows a clear
systematic difference between the Pierce \& Tully TF 
relation and the TF relation appropriate for the Zaritsky \& 
Rix data.  The TF residuals are correlated with the
velocity width, W$^{\rm i}$;  the correlation
is highly significant --- a Spearman rank test indicates a
likelihood of no correlation of P$_{\rm SR} = 3 \times 10^{-10}$.
Panel A shows the correlation between the asymmetry parameter
and the velocity width of the galaxy, which 
reflects a fundamental physical property of spiral galaxies ---
that low-mass galaxies are more asymmetric.
This correlation is also significant (P$_{\rm SR} = 0.018$).
Finally, Panel B shows the correlation Zaritsky \& Rix report
between the residual from the Pierce \& Tully (1992) $B$ TF 
relation and the stellar asymmetry of the $I$-band light. 
{\it The correlations in panels A and C give rise to
the apparent correlation in panel B}.  The solid line is the
least-squares fit to the data points in panel B; the dotted 
line coincident with the solid line is the
combination of the fits in panels A and C. The scatters match
well ($\sigma_{\rm A}=0.072$, $\sigma_{\rm C}=0.11$~km~s$^{-1}$, 
$\sigma_{\rm B}=0.073 \approx \sqrt{\sigma_{\rm A}^2 + 
({\rm b_A} \sigma_{\rm C})^2}$, where
${\rm b_A} =-0.15 {\rm\ km}^{-1} {\rm\ s}$ 
is the slope of the fit in panel A).  When we
remove the systematic TF slope problem by fitting the correlation
in panel C, the correlation between the TF residual and
the asymmetry disappears (P$_{\rm SR} = 0.70$).

A false correlation between the TF residuals and a third parameter
can arise only if the third parameter depends on $\eta$ and
M.  False correlations
can have two distinct (but related) causes:
systematic slope errors and selection effects.
The Zaritsky \& Rix example involves a 
systematic slope error.  With the proper slope,
this false correlation disappears. We address this concern
in our study
by testing for correlations using three different slopes:
the C97 slope (as measured by C97), -6.36,
the non-outlier pairs slope derived with the Willick (1994) technique,
-5.58, and the non-outlier pairs slope derived from direct least-squares
fitting, -4.75.  The last two rows of Table~\ref{tab:cmeasures}
show that using the steepest slope results in no correlation between
the residuals and M$_{\rm R}$ and that using the shallowest slope
results in no correlation between the residuals and $\eta$.

Selection effects are more complicated; they can give rise to
false correlations if the third parameter combines with
one or both of $\eta$ and M to influence sample selection.
For example, we analyze the $R$-band TF relation for a
$B$-selected sample of galaxies (from the Zwicky catalog).
Therefore, the $B-R$ color influences the selection of
intrinsically faint galaxies --- faint red galaxies are absent
from our sample.  Even without a
correlation between the TF residual and the color of a galaxy,
the sample selection results in a false correlation because
of the scatter in the TF relation.
``Underluminous'' red galaxies are underrepresented, and
the blue galaxies will seem preferentially underluminous.
In fact, we see no trend with color, suggesting that this effect is either
very small or, more likely, that excess
star formation cancels it.  This bias would damp any tendency for
blue galaxies to appear overluminous due to excess star formation.

These arguments suggest that false third parameter
correlations should have
some distinguishable characteristics.  A correlation may
be false if it doesn't persist for different slope
values. In addition,  a correlation 
is particularly suspicious if the third 
parameter is a strong function of both M and $\eta$.  In
this case, the correlation may still be real, especially
if it persists for multiple slope choices.
(Note, however, that if a third parameter/residual 
correlation {\it is} real, the third parameter should 
correlate with {\it either} M or $\eta$, by definition,
if the correlation is strong enough.)
Conversely, the failure to detect a correlation does
not mean that it is entirely absent.  Both large
scatter in the intrinsic TF distribution
and a large scatter in the relationship between
the third parameter and the residual associated
with the parameter can mask the true correlation.
We apply these arguments in the analysis below to 
distinguish between false and true correlations.

\subsection{Searching for Third-Parameter Dependence in the TF Relation}

In this section, we describe several 
statistics or measures, both discrete and continuous,
which we use to test for correlations between the TF residuals 
and the amount of tidal and kinematic distortion and
star formation in each galaxy.
Tables~\ref{tab:cmeasures} and \ref{tab:dmeasures} list the
continuous and discrete
statistics, respectively, which we describe below.
We exclude the 8 outliers from this analysis; they would
otherwise dominate the results. We list their parameters separately
in Table~\ref{tab:omeasures}.

\subsubsection{Measures of Distortion and Star Formation}

Objective measures of tidal or kinematic distortion are difficult
to define (Abraham et al. 1994; Odewahn et al. 1996; Windhorst et al. 1999;
Conselice, Bershady, \& Gallagher 2000;
Conselice, Bershady, \& Jangren 2000; Bershady, Jangren, \& Conselice 2000);
these measures 
are very sensitive to the exact surface brightness of the tidal features.  
Our statistics and
measures are not exhaustive descriptions of the galaxy;
they provide estimates of the amount of distortion
present.  

{\it Rotation Curve Distortion:} 
We consider two statistics which measure the amount of
rotation curve distortion.  
In Table~\ref{tab:cmeasures}, we 
list an objective measure, $\chi^2_{\rm rc}$, the $\chi^2$ per
degree of freedom of the fit to the (empirical) rotation curve
model of Courteau (1997).  
In addition, we divide the sample based on our estimate of the amount of
rotation curve distortion; 
we list this statistic in Table~\ref{tab:dmeasures}.
We classify the galaxies with
``normal'', ``marginal'', and ``distorted'' rotation
without {\it a priori}  knowledge of their placement on the TF
relation.

\begin{figure}
\centerline{\epsfysize=7.2in%
\epsffile{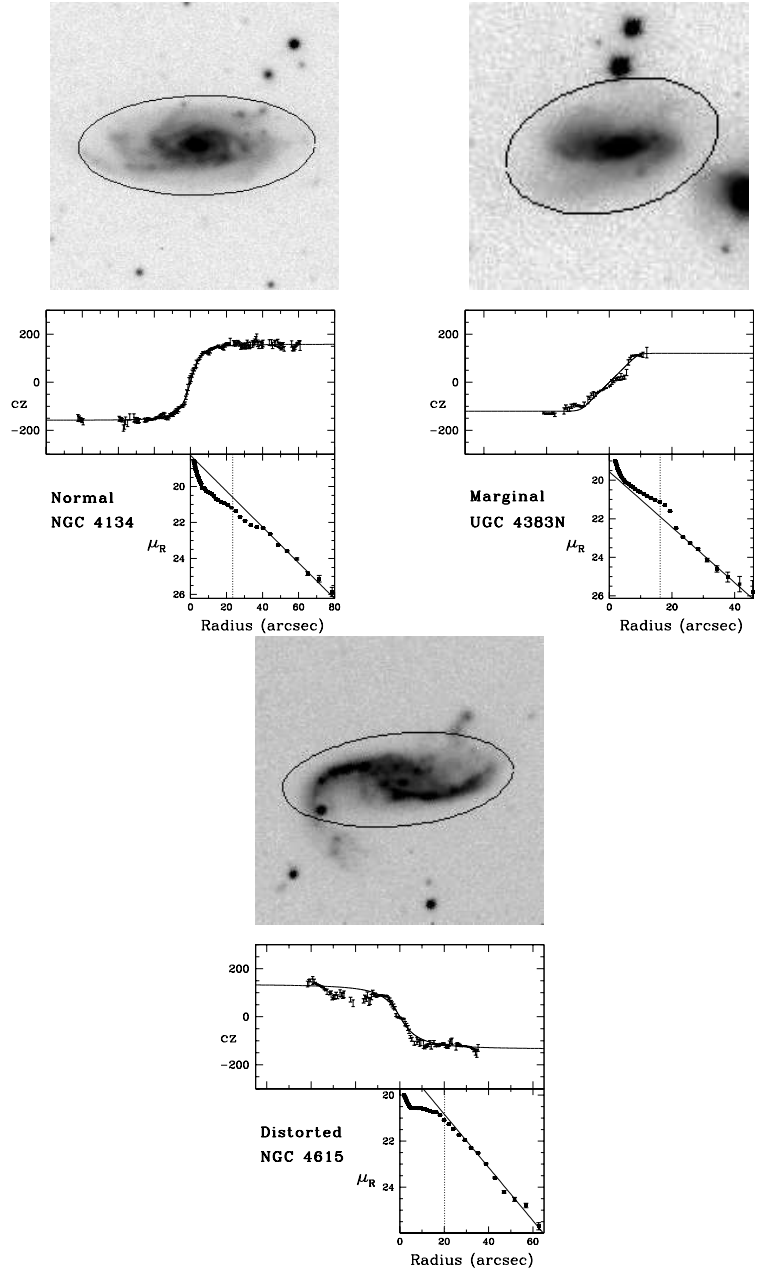}}
\caption{Examples of galaxies with ``normal'', ``marginal'', and
``distorted'' rotation curves.
For each galaxy,
the top panel is the $B$-band image, with the ellipse we use
to measure $\epsilon$.  The middle panel is the rotation curve,
on the same spatial scale as the image, and the bottom
panel is the $R$ surface brightness.  The vertical
line marks 2.15 R$_{\rm disk}$, the spatial position used to
measure V$_{\rm c}$.}
\label{fig:nmr}
\end{figure}

Fig.~\ref{fig:nmr} shows examples of the
three classes of rotation curves.  
The figure shows an image, surface brightness
profile and rotation curve of NGC~4134, 
which has a ``normal'' rotation curve.  
We also show a ``marginal'' case, UGC~4383N; 
the rotation curve barely turns over, 
providing essentially
a lower limit to V$_{\rm c}$.  
Finally, Fig.~\ref{fig:nmr} shows NGC~4615, 
which is significantly distorted,  even though
NGC~4615 is not an outlier to the TF relation.
Although this judgment-based measure gives a somewhat 
different picture of the effects of rotation curve distortion
from the more objective $\chi^2_{\rm rc}$, 
the mean reduced $\chi^2$ values 
from the fits to the C97 functions are 
4.1, 8.8, and 12.7, respectively,
for the 40 ``normal'' curves, 37 ``marginal'' curves,
and 13 ``distorted'' curves.

{\it Rotation Curve Truncation:} 
Our continuous measure of rotation curve truncation is
the maximum extent of the rotation curve (R$_{\rm max}$)
divided by the disk scale length, R$_{\rm disk}$.  Galaxies with large
values of R$_{\rm max}$/R$_{\rm disk}$ have well-sampled rotation curves.

{\it Morphological Distortion:}  We include 4 separate measures of
morphological distortion in the tables.  
The continuous morphological distortion measures are (1) 
R$_{\rm disk}$, the disk scale length, which can serve as a measure
of tidal stretching --- 
the few galaxies with the most prominent tidal tails 
have very large values of R$_{\rm disk}$; and
(2)  $\chi^2_{\rm phot}$, the $\chi^2$ per degree of freedom for the
linear fit to the surface brightness profile of the outer disk
(computed based on the ellipse fits, where the center, inclination,
and position angle are allowed to vary with radius).

We also include two discrete measures
of morphological distortion.  The first is a ``by-eye''
description of the disk as ``normal'',
``distorted'' or ``ambiguous''.  We compare only the
``normal'' and ``distorted'' cases in Table~\ref{tab:dmeasures}.
This judgment-based statistic 
is more sensitive than $\chi^2_{\rm phot}$ to relatively 
faint tidal features.  
Some galaxies which appear distorted have
low values of $\chi^2_{\rm phot}$; in most cases, the ``by-eye''
measure is effective.  
The final morphology measure is based on the
the measure of the ellipticity, $\epsilon$, 
of the galaxy's disk, and hence its inclination.  
We describe the
procedure for estimating the error, $\sigma_{\epsilon}$, in
Sec.~2.4.  Galaxies with stable, fit values of $\epsilon$ in their 
outer isophotes and $\sigma_{\epsilon} < 0.05$ are in the 
``well-defined $\epsilon$'' sample; galaxies with less stable values 
($\sigma_{\epsilon} > 0.05$) are in the ``poor $\epsilon$'' sample.
The ``no-fit $\epsilon$'' sample consists of galaxies for
which we can not fit ellipses at $\mu_{\rm R} \geq 24.5$;
it includes many of the most distorted galaxies (although it excludes the
few cases where the galaxy was fit by hand, typically classified
as ``good $\epsilon$'' galaxies).

{\it Star Formation:}  
In Table~\ref{tab:cmeasures} we list 4 measures of recent star formation:
the EW(H$\alpha$), which measures the most recent star formation, 
the color of the 
entire galaxy 
[corrected for reddening with the Courteau (1996) prescriptions], the
color of the center of the galaxy in the spectroscopic aperture of the BGK
observations (corrected for reddening based on the Balmer decrement), and
the strength of the new burst of star formation in $R$ (see Sec.~4.3).

{\it Star/Galaxy Overlap:} 
Finally, we divide the sample based on potential problems
with the photometry.  We consider galaxies which overlap their 
partners, and galaxies with substantial contamination from stars.
Two galaxies suffer from both, and are included in both subsamples.

\subsubsection{Searching for TF Residual Dependences}

Tables~\ref{tab:cmeasures} and \ref{tab:dmeasures} 
both describe the results of statistical tests for third-parameter
dependence in the TF relation.
For the continuous measures (Table~\ref{tab:cmeasures}) and
for three TF slopes, columns
3 -- 5  list 
the Spearman rank probabilities of no correlation between the TF residuals
and the third parameter.  In
columns 7 and 8, we list the
Spearman rank probabilities of no correlation between each 
third parameter and the TF parameters: the velocity width, $\eta$, 
and the total magnitude, M$_{\rm R}$, respectively.  
We list the correlations between the TF parameters and
the residuals in the last two rows.

The only possible correlations with TF residuals and
the continuous third parameters (columns 3 --
5 of upper rows, Table~\ref{tab:cmeasures})
are for $B-R$ color and R$_{\rm disk}$.  
Both color and R$_{\rm disk}$ correlate strongly with {\it both}
$\eta$ and M$_{\rm R}$, raising the possibility that the correlations
with the TF residuals
may result from the dependence of the residuals on
$\eta$ and M$_{\rm R}$.  The correlations
with TF residuals appear only for some slopes, supporting that conclusion.
R$_{\rm disk}$ correlates extremely strongly with M$_{\rm R}$;
it is suspicious that the strongest TF residual correlation
is for a slope of -4.75, which is the slope that results in the strongest
correlation between M$_{\rm R}$ and the residuals.
Similarly, $B-R$ is strongly correlated with $\eta$.
The TF residual/$B-R$ correlation is strong only for the
steep slope, which has the largest $\eta$ residuals.  
Therefore, we conclude that the apparent
correlations between residuals and both R$_{\rm disk}$ and
global color are probably false.

For the discrete measures (Table~\ref{tab:dmeasures}), columns
5 -- 7  list the Kolmogorov-Smirnov probabilities of the null hypothesis
that the distributions of TF residuals for subsamples broken down
based on the third parameters have been drawn from the same distributions.
In columns 8 and 9 we consider the distributions of $\eta$ and
M$_{\rm R}$ instead of the TF residuals. 
The only significant differences in TF residual distributions appear 
between galaxies with ``normal'' kinematics and galaxies with ``distorted''
kinematics.  For the least-squares slope, -4.75, the 9 galaxies with
``distorted'' rotation curves are ``overluminous'', with 
an average residual of -0.44 magnitudes using the least-squares
TF intercept, -20.98 magnitudes.
In addition, there are no differences in the $\eta$ distributions,
and at best only minor differences in the 
M$_{\rm R}$ distributions (possibly due
to the more fundamental TF residual dependence).
Therefore, we conclude that strong {\it kinematic} distortion
is a significant
predictor of TF residuals.   The required level of distortion appears in
only 9 of the 82 non-outlier galaxies, but  it appears in 4 of the
8 outliers (see Table~\ref{tab:omeasures}), 
further supporting our conclusion.  

\subsection{The Bottom Line: Luminosity Evolution Contributes
to the Slope Offset}

In Sec.~3.3, we measure a slope difference between the (non-outlier)
pair galaxies and the C97 sample.  
In Sec.~4, we argue that
star formation and kinematic distortion have the
largest effects on the TF properties of these
galaxies. 
Although we detect differences in the TF residuals based on
kinematic distortion, the 9 distorted galaxies 
do not cause the slope offset: 
if we remove these 9 points and repeat the Willick (1994) procedure,
using the Monte Carlo simulation to compute confidence intervals, 
the measured slope and offset 
stay nearly the same: $\Delta_{\rm TF} = -20.56 \pm 0.14$,
$\alpha_{\rm TF} = -5.71 \pm 0.47$, and 
$\sigma_{\rm TF} = 0.53 \pm 0.070$.  There is still a 
2.5$\sigma$ difference between the pairs slope and the C97
slope ($-7.03 \pm 0.26$).

Central star formation certainly contributes to the
flux of these galaxies, especially at the low-mass end (see,
e.g., column 5 of Table~\ref{tab:origins}).  
In Sec.~5.0, 
we argue that selection effects can damp the expected 
color dependence in the TF residuals caused by star formation.
A large scatter in the colors of the pre-interaction galaxies will
also damp this expected dependence.
Therefore, in spite of our failure to detect a color-residual dependence,
we conclude that star formation is probably a substantial 
contributor to the differences between the TF properties of our
sample and the C97 sample.

\begin{figure}
\centerline{\epsfysize=6.5in%
\epsffile{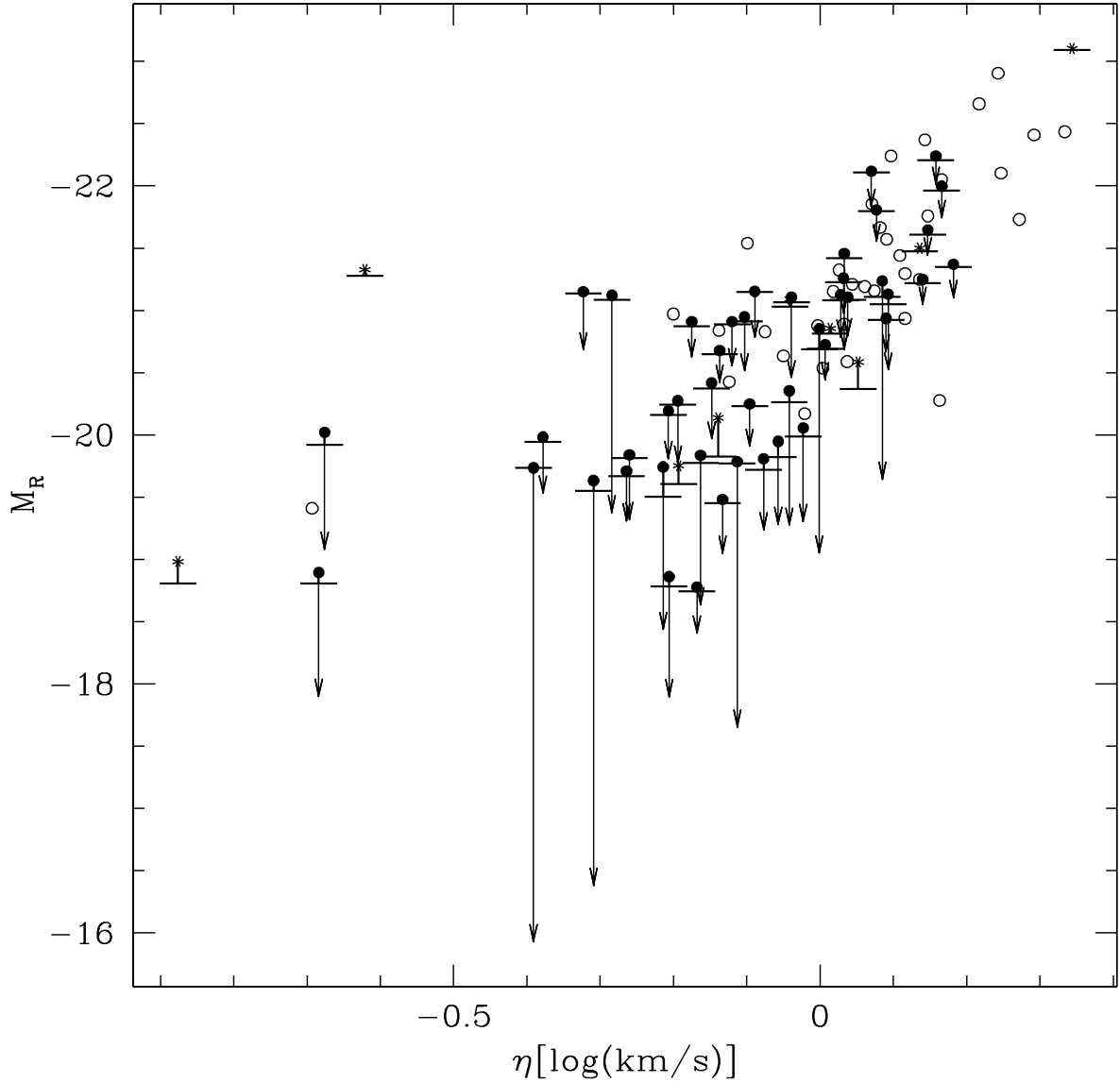}}
\caption{The possible effects of triggered star
formation.  We plot the TF relation
for the 89 pair galaxies with nuclear spectra.  
Assuming the Miller-Scalo IMF with
solar metallicity, a constant star formation rate,
and an old population with $B-R=1.5$, we plot galaxies with 
$0 < {\rm s_R} < 1$ as filled circles, galaxies with 
${\rm s_R = 1}$ as stars, and galaxies with little central 
star formation (hence no valid solution
for s$_{\rm R}$) as open circles, where s$_{\rm R}$ is the
new $R$ burst strength  
in the aperture of the nuclear spectrum.
For galaxies with s$_{\rm R} > 0$, the horizontal line
shows the lower-limit correction for the central burst only (i.e., the
magnitude with burst removed).  For galaxies with 
$0 < {\rm s_R} < 1$, the arrow shows the upper-limit correction,
assuming s$_{\rm R}$ is constant throughout.  Starred galaxies 
(with s$_{\rm R} = 1$) have no measured upper limit for the correction.
Note the change in y-axis scale from Fig.~\ref{fig:c97}.}
\label{fig:burstchanges}
\end{figure}

Table~\ref{tab:cmeasures} shows that the
burst strength, s$_{\rm R}$, correlates 
with velocity width (P$_{\rm SR} = 0.055$)
and luminosity (P$_{\rm SR} = 0.007$);
lower-mass galaxies have stronger bursts of star formation.
We measure s$_{\rm R}$ in only the central region of the
galaxy; the burst flux in this region provides a lower limit
to the total $R$-band magnitude change due to the burst:
$\Delta {\rm M_L} = 2.5 \log(1-{\rm f_{slit} s_R })$, 
where f$_{\rm slit}$ is the fraction of the
$R$ flux incident on the spectroscopic aperture (e.g.,
column 4 of Table~\ref{tab:origins}).
The burst is probably strongest in the central
region.  Therefore, the magnitude change
assuming a constant burst strength throughout the galaxy
(from disk star formation) is an upper limit:
$\Delta {\rm M_U} = 2.5 \log(1-{\rm s}_{\rm R})$, which is only
well-defined for s$_{\rm R} < 1$.  (Note that this quantity
is different from the quantity in
column 5 of Table~\ref{tab:origins}, for which we assume that
the galaxy lies on the C97 TF without the burst.)
For the Miller-Scalo IMF with constant star formation and
an assumed old population with color $B-R$=1.5, 
Fig.~\ref{fig:burstchanges} shows the range of possible
burst ``corrections'' to the TF properties of the sample.
For galaxies with $0 < {\rm s_R} < 1$ (solid points),
we show the lower-limit ``correction'' for the central burst only
(horizontal line) and the upper-limit ``correction'' assuming
a constant burst strength (arrow).  
Most of the high-mass
galaxies are firmly anchored (but two high-mass exceptions have
s$_{\rm R} = 1$, the starred points). However, star formation
can affect the low-mass galaxies substantially.
The range of possible corrections for the low-mass galaxies
is large, allowing for substantial steepening of the
slope with subtraction of the recent star formation.

We estimate the upper limit of the effects of 
triggered star formation on the low-mass galaxies
empirically from our TF results,
assuming the differences between the C97 TF and the
non-outlier pair TF at the low-mass end
are entirely due to star formation.
The relations intersect at $\eta_{\star} = -0.155$; for
$\eta \leq \eta_{\star}$, a pair galaxy 
luminosity is greater than that of a typical C97 galaxy.
The increase in flux is equal to 
$\delta M = 1.45(\eta_{\star} - \eta)$, a boost of
0.35 magnitudes, or 39\% in flux, at the low-mass end, $\eta = -0.4$.
These values correspond to {\it average} bursts of
up to -17.1 magnitudes in $R$. This burst size is very sensitive
to the measured zero-point offset.  If the true zero-point of
the C97 relation is 0.1 magnitudes fainter (see Sec.~3.3), 
we expect average bursts of up to ${\rm M_R = -17.7}$.

Bursts in this size range would be undetectable
at the high-mass end of the TF relation (see Fig.~\ref{fig:c97bursts}).
The Leitherer et al. (1999) 
models (with solar metallicity)
show that constant star formation rates of 
$\sim$ 0.5 -- 2.6 M$_{\sun}$~yr$^{-1}$ for 10$^7$ years or,
much more likely,
$\sim$ 0.2 -- 0.4 M$_{\sun}$~yr$^{-1}$  for 10$^8$ years 
(see Barton et al. 2000a), 
can form enough new stars to account for M$_{\rm burst} = -17.1$, 
where the range of values results from the
the range of stellar initial mass functions in the models.
With a 0.1-magnitude increase in the measured
C97 $R$-band zero point (due to $r$-to-$R$ conversion errors
or correction for diameter limit bias),
these star formation rates can nearly double and still be consistent
with the data.  
Only a fraction of the galaxies in our sample are undergoing
triggered star formation; thus the actual change in slope
due to star formation probably results from stronger bursts in 
fewer than half of the galaxies.  
Our bursts of $\lesssim 0.5$ magnitudes
are marginally significant and the constraints this technique
places on their strengths are only statistical in nature.

\section{Detecting Star Formation at High Redshift}

The cosmological assembly of galaxies at very high redshift 
and the subsequent variations of the 
star formation rate as a function of redshift
are constraints for galaxy evolution 
(e.g., Kauffman \& White 1993; Somerville \& Primack 1999;
Kolatt et al. 1999).  
The TF relation provides
an important tool for measuring these effects.
Hierarchical models
predict an increased rate of mergers and interactions
at high redshift,  borne out by observations of an
increased incidence of irregular
morphology at 
high redshift (Abraham et al. 1994; van den Bergh et al. 1996;
Odewahn et al. 1996).
Our study focuses on the best objectively-selected
candidates for evolving galaxies at the current epoch: galaxies
in pairs.  Although minor mergers are absent from our
sample, the processes by which minor mergers trigger
gas infall and star formation (e.g., Mihos \& Hernquist
1994) are qualitatively similar to the triggering
processes in major interactions.  Our results
are broadly applicable as a baseline for TF studies of
evolving galaxies at high redshift.

The potential for TF outliers and for greater
scatter increases the amount of data necessary to 
detect an elevation in the star formation rate at 
high redshift.  Some outliers can be eliminated
based on kinematic distortion observable even at high redshift
(e.g., NGC~7235B in Fig.~\ref{fig:out1}).  However,
most are more difficult to prune on this basis.
Any high-redshift TF study should include enough
data in each redshift range to define a locus for typical
galaxies and to eliminate the outliers.  

Approximately 90\% of the pair galaxies in our sample
lie reasonably close to the TF relation of C97.  Thus, 
even if the slope and offset deviations in our sample
are due entirely to ``parameter misinterpretation errors'' 
(which they are not), tidal distortions are relatively 
ineffectual.  High-redshift samples the size
of this study should be able to detect luminosity
evolution $\gtrsim 0.5$ magnitudes or better at $R$, 
even if the samples include interacting galaxies.

Although longer rest-frame wavelengths like $R$ are
reliable due to small scatter and less sensitivity to 
very recent star formation, 
observational considerations frequently make 
studies of the rest-frame $B$ TF properties more
convenient.  We apply the Willick (1994) technique to
the $B$-band data [corrected for extinction with
the Courteau (1996) prescriptions, with corrections 4/2.2 times larger
in $B$].  
With no outlier clipping, we find TF parameters
(${\rm \Delta_{TF}, \alpha_{TF}, \sigma_{TF}}$) = 
(-18.94, -3.02, 1.01).  
Clipping with the
Tully \& Pierce (2000) $B$-band slope of -7.27 yields
8 outliers; 7 are the same as the $R$-band outliers,
including all of the low-mass outliers.
The ``underluminous'' UGC~4744 is not a $B$ outlier,
and the very blue Northern half of NGC~3991 is,
with (corrected) $B-R=0.33$. 
With these 8 outliers removed, we find $B$-band
TF parameters (${\rm \Delta_{TF}, \alpha_{TF}, \sigma_{TF}}$) = 
(-19.33, -4.88, 0.68).  

The set of outliers identified in a high redshift study
should contain some of the most interesting objects,
presumably including anomalously narrow emission line galaxies
(like the CNELGs).  The relatively high
incidence of these objects in our pair
sample occurs because of close galaxy-galaxy passes,
which can drive a large fraction of the disk gas 
into the galaxy center.  The stars 
formed may contribute to the bulge.
Thus, TF studies where sample pruning is minimized 
may provide a powerful means of selecting galaxies evolving
along the Hubble sequence.

\section{Conclusion}

We use optical emission line rotation curves to 
investigate the $R$-band Tully-Fisher properties of a sample of 
90 spiral galaxies in close pairs.
The sample includes a range of 
luminosities, morphological types, and degrees of tidal
distortion.  
With the exception of
eight distinct $\sim$3$\sigma$ outliers,
the galaxies follow the Tully-Fisher relation
remarkably well, 
even in the presence of striking
tidal distortion.
Although most of the outliers
show signs of strong star formation during the
last $\sim$100 -- 300 Myr, 
gasdynamical effects are probably
the dominant cause of their
anomalous Tully-Fisher properties.  Strong effects from
morphological distortion and/or dissipationless kinematic
distortion are rare.

Four outliers with small emission line widths have 
very centrally concentrated emission
and truncated rotation curves.
Recent gas infall after a close
galaxy-galaxy pass or a minor merger 
can trigger this emission and
can enhance the bulges of spiral galaxies.
These four galaxies may be local 
counterparts to the compact, blue galaxies observed
at intermediate redshift, including the 
compact narrow emission line galaxies.

The remaining galaxies have only a small zero-point
offset from the Courteau (1997) TF relation for more isolated 
galaxies, but the pairs have a shallower
slope (2.6$\sigma$ significance) and a 25\%  
larger scatter. 
We argue that triggered star formation is one significant
contributor to the slope difference.  
We characterize the non-outlier sample with measures of
distortion and star formation
to search for third parameter dependence in the
residuals of the TF relation.
Severe kinematic distortion is the only
significant predictor of TF residuals, although this 
distortion is not responsible for the slope difference.

Our results have several implications for TF studies at
moderate or high redshift:
\begin{enumerate}

\item Morphological distortion alone rarely results in
strong effects on the TF properties of galaxies.  
Gasdynamical effects on the rotation curve are much
more common.  These effects together give rise to a small set
of outliers ($\sim$10\% of our sample).

\item Outliers due to interaction-induced distortion
are easily removed by sigma clipping.  
This clipping provides a method for identifying
galaxies with concentrated emission in an epoch
of bulge enhancement, such as the compact, narrow
emission line galaxies.

\item Luminous galaxies require prohibitively large
bursts of star formation to move off the TF relation;
at least locally, bursts of star formation will
affect only low-mass galaxies.
Therefore, the slope of the
TF relation, while difficult to measure, is 
at least as fundamental for quantifying luminosity
evolution as the zero-point offset.

\end{enumerate}

We conclude that detection of moderate 
($\gtrsim 0.5$ magnitudes in rest-frame $R$) luminosity evolution
should be possible with
high-redshift samples the size of this 90-galaxy
study, even in the presence of some tidal distortion.

\begin{acknowledgements}
 
We thank Norman Grogin for the 
use of his additions to the  GALPHOT package, and for other software 
and useful suggestions. We thank Sheila Kannappan for 
insightful comments on linewidth measurement techniques,
Daniel Koranyi and Jan Kleyna for 
useful comments and software relating to galaxy coordinate systems 
and photometric calibrations, and Jeffrey Sutherland for helping
to analyze the simulations.
This research is supported in part by the Smithsonian Institution.
We are grateful to the Caltech Center for Advanced Computing
Research and NASA Offices of Space Sciences, Aeronautics, and Mission
to Planet Earth for providing computing resources.
 
\end{acknowledgements}

\clearpage

\clearpage

\clearpage

\clearpage

\clearpage

\begin{deluxetable}{ll}
\tablenum{1}
\tablewidth{0pc}
\tablecolumns{2}
\tablecaption{TF Parameters}
\tablehead{ \colhead{Parameter} & \colhead{Description} }
\startdata
m$_{\rm R} $ & total $R$ magnitude\nl
m$_{\rm B} $ & total $B$ magnitude\nl
$\epsilon$ & disk ellipticity\nl
R$_{\rm disk}$ & radial disk scale length\nl
$m$ & slit misalignment angle (in the plane of the sky)\nl
V$_{2.2}$ & velocity width at 2.15 $\times$ R$_{\rm disk}$ (C97)\nl
\enddata
\label{tab:measuredparams}
\end{deluxetable}

\begin{deluxetable}{ccccccc}
\tablenum{2}
\tablewidth{0pc}
\tablecolumns{7}
\tablecaption{TF Properties}
\tablehead{
                        & \multicolumn{3}{c}{All Curves} & \multicolumn{3}{c}{$R$-Band Outliers Removed} \\ 
 \colhead{Reference}    & & \colhead{$\Delta_{\rm TF}$} & \colhead{Scatter}  &  & \colhead{$\Delta_{\rm TF}$} & \colhead{Scatter} \\
 \colhead{Distribution} & \colhead{N$_{\rm gal}$} & \colhead{(mag.)} & \colhead{(mag.)} & \colhead{N$_{\rm gal}$} & \colhead{(mag.)} & \colhead{(mag.)} }
\startdata
Courteau (1997) $r$        & 90 & -0.06 $\pm$ 0.10 & 1.00 & 82 &  0.14 $\pm$ 0.06 & 0.55 \nl
Tully \& Pierce (2000) $R$ & 89 & -0.46 $\pm$ 0.11 & 1.01 & 81 & -0.26 $\pm$ 0.07 & 0.60 \nl
Tully \& Pierce (2000) $B$ & 89 & -0.50 $\pm$ 0.11 & 1.04 & 81 & -0.30 $\pm$ 0.07 & 0.66 \nl
Pierce \& Tully (1992) $R$ & 90 & -0.87 $\pm$ 0.13 & 1.26 & 82 & -0.61 $\pm$ 0.08 & 0.75 \nl 
Pierce \& Tully (1992) $B$ & 90 & -0.75 $\pm$ 0.14 & 1.28 & 82 & -0.50 $\pm$ 0.09 & 0.81 \nl
\tablecomments{TF Statistics: (1) Reference distribution, 
(2) number of galaxies in the sample,
(3) offset from the reference distribution, based on a fit of
the TF relation with the slope fixed by the reference distribution; a negative
offset indicates that the galaxies appear {\it overluminous} with respect
to the reference distribution, and (4) scatter from the fit in (3).
One galaxy is missing from the
TP00 fits due to the different prescription 
for computing the inclination angle.
(5) -- (7) are the same as (2) -- (4), for the distributions without the 
8 $R$-band outliers. }
\enddata
\label{tab:tf}
\end{deluxetable}

\begin{deluxetable}{lrrrrr}
\tablenum{3}
\tablewidth{0pc}
\tablecolumns{6}
\tablecaption{Relations Between m$_{\rm Zw}$ and the TF Parameters}
\tablehead{    &                         & \colhead{ a }    & \colhead{ b }          & \colhead{ c} & \colhead{$\sigma_{\rm m,Zw}$} \\
\colhead{Data} & \colhead{N$_{\rm gal}$} & \colhead{(mag.)} & \colhead{(mag./dex) }  & \colhead{(mag./mag.)} & \colhead{(mag.)} }
\startdata
C97 ``field'' Sample & 279 & -1.370 & 1.711 & 1.187 & 0.421 \nl
Pairs & 90 & 0.797 & 1.114 & 1.070 & 0.546 \nl
\enddata
\tablecomments{Fits to the relation m$_{\rm Zw} = {\rm a} + 
{\rm b} \eta + $~cm, where m$_{\rm Zw}$ is the Zwicky magnitude, 
$\eta$ is the velocity width parameter, and m is the apparent
magnitude ($R$-band for the pairs data, Gunn $r$ for the C97 data),
using the method of Willick (1994):
(1) data set, (2) number of points, (3) -- (5) results, (6) scatter.}
\label{tab:mzrel}
\end{deluxetable}

\begin{deluxetable}{lrrrr}
\tablenum{4}
\tablewidth{0pc}
\tablecolumns{5}
\tablecaption{TF Relations}
\tablehead{    &                         & \colhead{$\Delta_{\rm TF}$}& \colhead{$\alpha_{\rm TF}$} & \colhead{$\sigma_{\rm TF}$} \\
\colhead{Data} & \colhead{N$_{\rm gal}$} & \colhead{(mag.)}           & \colhead{(mag./dex)}        & \colhead{(mag.)} }
\startdata
C97 Sample   & 279 & -20.83 $\pm$ 0.05 & -7.03 $\pm$ 0.26 & 0.45 $\pm$ 0.03 \nl
Pairs              & 90  & -20.12 $\pm$ 0.40 & -3.77 $\pm$ 0.55 & 0.94 $\pm$ 0.15 \nl 
Pairs, no outliers & 82  & -20.61 $\pm$ 0.15 & -5.58 $\pm$ 0.50 & 0.56 $\pm$ 0.07 \nl
\enddata
\tablecomments{Fits to the TF relation 
M$_{\rm R} = \Delta_{\rm TF} + \alpha_{\rm TF} \eta$, where
$\eta$ is the velocity width parameter, and M is the magnitude 
($R$-band for the pairs data, converted to $R$ from Gunn $r$ for the C97 data):
(1) data set, (2) number of points, (3) -- (4) results, (5) scatter.  The errors are the
68.3\% confidence levels from the Monte Carlo simulation for the
individual, not joint, parameters (e.g., 68.3\% of all the simulations
for the C97 sample had a computed offset within -20.83 $\pm$ 0.05).}
\label{tab:tfparams}
\end{deluxetable}

\begin{deluxetable}{llrrrrrrrrr}
\tablenum{5}
\tablewidth{0pc}
\tablecolumns{11}
\scriptsize
\tablecaption{Outlier Properties}
\tablehead{
                  &                 &                        &                  &                       &                       & \colhead{EW(H$\alpha$)} & \colhead{EW(H$\delta$)} &                  &  \colhead{R$_{\rm e,R}$} & \colhead{R$_{\rm e,B}$} \\
 \multicolumn{2}{c}{Galaxy/Label} & \colhead{z$_{\rm LG}$} & \colhead{$\eta$} & \colhead{M$_{\rm R}$} & \colhead{M$_{\rm B}$} & \colhead{(\AA)}         & \colhead{(\AA)}         & \colhead{$B-R$} & \colhead{(h$^{-1}$ kpc)} & \colhead{(h$^{-1}$ kpc)} }
\startdata
UGC 4744     & A & 0.0078 &  0.16 & -20.27 & -18.83 &  0.0 $\pm$ 1.0 & -1.3 $\pm$ 0.3 & 1.44  & 1.8 & 1.8\nl
UGC 312E      & B & 0.0152 & -0.28 & -21.12 & -20.53 & 71.0 $\pm$ 1.4 &  0.8 $\pm$ 0.8 & 0.59  & 5.2 & 5.5\nl
NGC 7253B    & C & 0.0162 & -0.32 & -21.14 & -20.14 & 15.4 $\pm$ 1.2 &  0.0 $\pm$ 0.7 & 1.00  &10.4 &10.6\nl
UGC 6944     & D & 0.0111 & -0.62 & -21.33 & -20.68 & 79.3 $\pm$ 1.6 &  0.0 $\pm$ 0.3 & 0.65  & 5.0 & 5.0\nl
UGC 7085W    & E & 0.0210 & -0.68 & -20.03 & -19.30 & 21.4 $\pm$ 1.4 & -3.8 $\pm$ 0.3 & 0.74  &  2.3& 2.1\nl
NGC 2719A    & F & 0.0103 & -0.68 & -18.89 & -18.47 &100.2 $\pm$ 3.0 &  4.2 $\pm$ 0.3 & 0.42  &  1.2& 1.2\nl
UGC 8919N    & G & 0.0095 & -0.69 & -19.40 & -18.31 &  7.1 $\pm$ 1.0 & -3.9 $\pm$ 0.3 & 1.09  &  1.9& 1.8\nl
CGCG 132-062 & H & 0.0094 & -0.88 & -18.99 & -18.02 & 62.0 $\pm$ 1.1 &  0.0 $\pm$ 0.1 & 0.97  &  1.2& 1.0\nl
\enddata
\tablecomments{Properties of the TF Outliers: (1) Name, (2) label for
Fig.~\ref{fig:c97}, (3) redshift, corrected for motion in the local
group, (4) velocity width parameter, (5) M$_{\rm R}$ (corrected for extinction
with the C97 prescriptions), (6) M$_{\rm B}$ (corrected), (7) equivalent width 
of H$\alpha$, (8) equivalent
with of H$\delta$ (a negative value denotes absorption), (9) $B-R$, (10) R half-light radius, and (11) B half-light radius.}
\label{tab:outliers}
\end{deluxetable}

\begin{deluxetable}{lrcrccl}
\tablenum{6}
\tablewidth{0pc}
\tablecolumns{7}
\tablecaption{More Outlier Properties}
\tablehead{
\colhead{} & \colhead{} & \colhead{M$_{\rm R, burst, slit}$} & \colhead{} & \colhead{Expected residual} & \colhead{TF residual} & \colhead{} \\
\colhead{Galaxy} & \colhead{s$_{\rm R}$} & \colhead{(mag.)} & \colhead{f$_{\rm slit}$} & \colhead{(mag.)} & \colhead{(mag.)} & \colhead{Comment} }
\startdata
UGC 4744     & ---   & ---    & ---     & ---& +1.89 & ambiguous\nl
UGC 312E      & 0.80 & -17.3 &  3.8\% & -0.16 & -1.81 & ambiguous\nl
NGC 7253B    & 0.35 & -16.4 &  3.7\% & -0.09 & -2.07 & poorly-sampled curve\nl
UGC 6944     & 1.00 & -17.9 &  4.3\% & -1.17 & -4.16 & tidal elongation\nl
UGC 7085W    & 0.58 & -17.4 & 15.5\% & -1.08 & -3.21 & truncated curve\nl
NGC 2719A    & 0.60 & -16.1 & 13.1\% & -0.47 & -2.12 & truncated curve\nl
UGC 8919N    & ---   & ---    & ---     & ---& -2.69 & truncated curve\nl
CGCG 132-062 & 1.00 & -16.9 & 14.9\% & -1.63 & -3.44 & truncated curve\nl
\enddata
\tablecomments{Origins of the TF Outliers: (1) Name, (2) burst strength in $R$, or the
fraction of $R$ flux in the
slit of the BGK spectroscopic aperture that originates from the new burst (computed
with the technique of Fig.~\ref{fig:deburst}), (3) $R$ magnitude of this 
central star formation in the
spectroscopic aperture, (4) fraction of the total $R$ flux of the galaxy 
that is in the spectroscopic aperture, (5) expected TF shift from the new 
flux in the central aperture (assuming an initial luminosity appropriate for
its $\eta$ and the C97 TF relation), (6) actual residual from the C97
TF relation, (7) our best guess as to the primary physical origin of the TF residual
(see Secs.~4.2 -- 4.4).
Two galaxies (UGC~4774 and UGC~8919N) are outside the contour
boundaries of Fig.~\ref{fig:deburst}; we have no burst strength
solutions for these galaxies. }
\label{tab:origins}
\end{deluxetable}

\begin{deluxetable}{llccccc}
\tablenum{7}
\small
\tablewidth{0pc}
\tablecolumns{7}
\tablecaption{Pair TF Residuals vs. Continuous Third Parameters, Spearman Rank Correlation Tests}
\tablehead{
& & \multicolumn{5}{c}{Spearman Rank Probabilities} \\
&  & \colhead{C97} & \colhead{unbiased} & \colhead{least-squares} & \colhead{$\eta$} & \colhead{M$_{\rm R}$} \\
\colhead{Test} & \colhead{Measure} & \colhead{(-6.36)} & \colhead{(-5.58)} & \colhead{(-4.75)} & \colhead{log(km/s)} & \colhead{(mag.)} }
\startdata
Kinematic            & $\chi^2_{\rm rc}$                 & 0.101 & 0.177 & 0.360 & 0.017 & 0.084 \nl
\hspace{0.1in} distortion       & & & & & & \nl
Curve truncation      & R$_{\rm max}$/R$_{\rm disk}$     & 0.273 & 0.662 & 0.995 & 0.170 & 0.241 \nl
Tidal stretching      & R$_{\rm disk}$ (h$^{-1}$ kpc)    & 0.233 & 0.010 & $2.26 \times 10^{-5}$ & $2.02 \times 10^{-10}$ & $4.28\times 10^{-18}$ \nl
Morphological         & $\chi^2_{\rm phot}$              & 0.672 & 0.891 & 0.956 & 0.466 & 0.632 \nl
\hspace{0.1in} distortion   & &  &  &  &  &  \nl
Star formation        & EW(H$\alpha$) \tablenotemark{a}  & 0.274 & 0.783 & 0.407 & $6.80 \times 10^{-6}$ & $3.89 \times 10^{-6}$ \nl
Star formation        & $B-R$, galaxy                    & 0.023 & 0.198 & 0.998 & $8.24 \times 10^{-9}$ & $6.36 \times 10^{-7}$ \nl
Star formation        & $B-R$, center \tablenotemark{a}  & 0.683 & 0.709 & 0.152 & $1.86\times10^{-5}$ & $3.63\times10^{-6}$ \nl
Burst strength        & s$_{\rm R}$ \tablenotemark{b}    & 0.500 & 0.248 & 0.149 & 0.055 & $6.82\times10^{-3}$ \nl
\tableline
TF parameters         & $\eta$                           & $3.14 \times 10^{-4}$ & 0.049 & 0.95 &  &  \nl
                      & M$_{\rm R}$                      & 0.425 & 0.016 & $5.57 \times 10^{-6}$ &  &  \nl
\enddata
\tablenotetext{a}{Includes only the 89 galaxies with separate nuclear spectra.}
\tablenotetext{b}{Includes only the 48 galaxies with valid solutions.}
\tablecomments{Spearman rank probabilities of no correlation with third
parameters: (1) quantity that parameter tests, (2) parameter 
(see Sec.~5.2), (3) -- (5) P$_{\rm SR}$ of no correlations with 
the TF residuals, assuming the C97 slope, -6.36, the slope from the 
Willick (1994) technique,
-5.58, and the straight linear least-squares slope for the pairs 
(with no outliers), -4.75 respectively, and
(6) -- (7) P$_{\rm SR}$ of no correlations 
with the TF parameters, $\eta$ and M$_{\rm R}$.}
\label{tab:cmeasures}
\end{deluxetable}

\begin{deluxetable}{llccccccc}
\tablenum{8}
\small
\tablewidth{0pc}
\tablecolumns{9}
\tablecaption{TF Residual -- Discrete Third Parameter Kolmogorov-Smirnov Tests}
\tablehead{
& & & & \multicolumn{5}{c}{Kolmogorov-Smirnov Probabilities} \\
& & & & \colhead{C97} & \colhead{unbiased} & \colhead{least-squares} & \colhead{$\eta$} & \colhead{M$_{\rm R}$} \\
\colhead{Test} & \colhead{Measure} & \colhead{N$_1$} & \colhead{N$_2$} & \colhead{(-6.36)} & \colhead{(-5.58)} & \colhead{(-4.75)} & \colhead{log(km/s)} & \colhead{(mag.)} }
\startdata
Kinematic  & ``Normal'' vs. & 39 & 34 & 0.810 & 0.681 & 0.934 & 0.110 & 0.229\nl
\hspace{0.1in} distortion & ``Marginal''   & & & & & & \nl
           & ``Normal'' vs. & 39 &  9 & 0.003 & 0.005 & 0.003 & 0.848 & 0.100\nl
 & ``Distorted''  & & & & & & \nl
\tableline
Morphological   & ``Normal'' vs  & 31 & 39 & 0.359 & 0.259 & 0.085 & 0.953 & 0.125\nl
\hspace{0.1in} distortion      & ``Distorted''   & & & & & & \nl
$\epsilon$                & well-determined $\epsilon$  vs. & 60 & 22 & 0.774 & 0.659 & 0.606 & 0.185 & 0.069\nl
 & poor, no-fit  & & & & & & \nl
\tableline
Overlap        & unobstructed vs.  & 37 & 31 & 0.863 & 0.908 & 0.548 & 0.201 & 0.246\nl
 & overlapping galaxies & & & & & & \nl
                           & unobstructed vs.  & 37 & 12 & 0.916 & 0.804 & 0.901 & 0.489 & 0.375\nl
 & overlapping star(s) & & & & & & \nl
\tableline
\enddata
\tablecomments{Kolmogorov-Smirnov probabilities for the distributions
of TF residuals for subsamples separated based on discrete
third parameters: (1) quantity that parameter measures, 
(2) distributions compared, (3) number of galaxies in first
set, (4) number of galaxies in second set, (5) -- (7) P$_{\rm KS}$ for
the distributions of TF residuals, measured assuming 
the C97 slope, -6.36, the slope from the Willick (1994) technique,
-5.58, and the straight linear least-squares slope for the pairs 
(with no outliers), -4.75 respectively, and 
(8) -- (9) P$_{\rm KS}$ for the distributions of 
$\eta$ and M$_{\rm R}$.}
\label{tab:dmeasures}
\end{deluxetable}

\begin{deluxetable}{lrrrrrrrr}
\tablenum{9}
\small
\tablewidth{0pc}
\tablecolumns{9}
\tablecaption{Outlier Parameters}
\tablehead{
\colhead{} &
\colhead{UGC} &
\colhead{UGC} &
\colhead{NGC} &
\colhead{UGC} &
\colhead{UGC} &
\colhead{NGC} &
\colhead{UGC} &
\colhead{CGCG} \\
\colhead{Parameter} &
\colhead{4744} &
\colhead{312E} &
\colhead{7253B} &
\colhead{6944} &
\colhead{7085W} &
\colhead{2719A} &
\colhead{8919N} &
\colhead{132-062} }
\startdata
$\chi^2_{\rm rc}$              & 2.9 &7.1  & 11.2&12.4 & 5.5 &1.4 & 3.0& 3.0\nl
R$_{\rm max}$/R$_{\rm disk}$   & 0.3 &5.2  & 3.5 &4.5  &1.8  &2.9 & 0.2& 1.4\nl
R$_{\rm disk}$ (h$^{-1}$ kpc)  & 8.2 &2.4  & 1.8 &2.2  &1.8  &1.0 & 3.6& 0.9\nl
$\chi^2_{\rm phot}$            & 0.2 &18.1 & 81.1&1.5  &0.2  &0.2 & 0.1& 1.8\nl
EW(H$\alpha$)                  & 0.0 &71.0 & 15.4&79.3 &21.4 &100.2&7.1& 62.0\nl
$B-R$, galaxy                  & 1.4&0.6 & 1.0 &0.6 &0.7 &0.4& 1.1& 1.0\nl
$B-R$, center                  & 1.2&0.4 & 0.9 &0.2 &0.7 &0.7& 0.6& -0.1\nl
s$_{\rm R}$                    & --  & 0.8& 0.4 &1.0 &0.6 &0.6 & --& 1.0\nl
Rotation curve \tablenotemark{a}             & N  & D & D & D & D & M & M & M\nl
Morphology \tablenotemark{b}                 & D  & D & D & D & N & N & N & N\nl
$\epsilon$ measurement \tablenotemark{c}  & F  & P & P & W & W & F & W & P\nl
Overlapping? \tablenotemark{d}               & S/G& N & G & S & S & G & G & S\nl
\enddata
\tablenotetext{a}{N=normal, M=marginal, D=distorted}
\tablenotetext{b}{N=normal, U=unknown, D=distorted}
\tablenotetext{c}{W=well-determined, P=poor fit, F=not fit}
\tablenotetext{d}{N=normal, S=star(s) overlapping, G=galaxies overlap}
\tablecomments{``Third parameter'' values for the 8 outliers.  See the
parameter descriptions in Sec.~5.1.}
\label{tab:omeasures}
\end{deluxetable}

\end{document}